\documentclass{aa}
\usepackage{txfonts}

\usepackage[pdftex]{graphicx}

\DeclareGraphicsExtensions{.jpg}

\usepackage{amssymb}
\usepackage{natbib}
\bibpunct{(}{)}{;}{a}{}{,} 

\newcommand{\va}{\vec{a}}
\newcommand{\vn}{\vec{n}}
\newcommand{\vx}{\vec{x}}

\newcommand{\vs}{\vec{s}}
\newcommand{\vp}{\vec{p}}
\newcommand{\tR}{\tens{R}}

\title{A full sky, low foreground, high resolution CMB map from WMAP}

  \author{J. Delabrouille
         \inst{1}
         \and
         J.-F. Cardoso\inst{1,2}
         \and
         M. Le Jeune\inst{1}
         \and
         M. Betoule \inst{1}
         \and
         G. Fay \inst{1,3}
         \and
         F. Guilloux \inst{1,4,5}
         }

  \offprints{delabrouille@apc.univ-paris7.fr}

  \institute{
    Laboratoire AstroParticule et Cosmologie (APC), CNRS: UMR 7164,
    Universit\'e Paris Diderot - Paris 7,\\
    10, rue A. Domon et L. Duquet, 75205 Paris Cedex 13, France\\
    \and
    Laboratoire de Traitement et Communication de l'Information (LTCI), CNRS: UMR 5141, T\'el\'ecom ParisTech,\\
    46 rue Barrault, 75634 Paris Cedex, France\\
    \and
    Laboratoire Paul Painlev\'e, CNRS: UMR 8524, Universit\'e Lille 1,\\
    59 655 Villeneuve d'Ascq Cedex, France\\
    \and
    Mod\'elisation Al\'eatoire de Paris X (MODAL'X), Universit\'e Paris Ouest --
    Nanterre-la-D\'efense,\\
    200 avenue de la R\'epublique, 92001 Nanterre Cedex, France\\
    \and
    Laboratoire de Probabilit\'es et Mod\`eles Al\'eatoires (LPMA),
    CNRS: UMR 7599, Universit\'e Paris Diderot - Paris 7,\\
    175 rue du Chevaleret, 75013 Paris, France\\
   }

  \date{\today}

\def\lm{{\ell m}}
\def\mcJ{\mathcal J}
\begin{document}

\abstract{
The WMAP satellite has made available high quality maps of the sky in five frequency bands ranging from 22 to 94 GHz, with the main scientific objective of studying the anisotropies of the Cosmic Microwave Background (CMB). These maps, however, contain a mixture of emissions from various astrophysical origins, superimposed on CMB emission.}{The objective of the present work is to make a high resolution CMB map in which contamination by such galactic and extra-galactic foregrounds, as well as by instrumental noise, is as low as possible.}{The method used is an implementation of a constrained linear combination of the channels with minimum error variance, and of Wiener filtering, on a frame of spherical wavelets called needlets, allowing localised filtering in both pixel space and harmonic space.}{We obtain a low contamination low noise CMB map at the resolution of the WMAP W channel, which can be used for a range of scientific studies. We obtain also a Wiener-filtered version with minimal integrated error.}{The resulting CMB maps offer significantly better rejection of galactic foregrounds than previous CMB maps from WMAP data. They can be considered as the most precise full--sky CMB temperature maps to-date.}


\maketitle

\section{Introduction}

The WMAP satellite is one of the most successful experiments dedicated to mapping the Cosmic Microwave Background (CMB). The all-sky maps obtained in the WMAP five frequency bands, in temperature and polarisation, offer the best data set to-date for making a sensitive all-sky map of the CMB anisotropies.

The CMB, however, is not the only source of emission at WMAP frequencies. Diffuse galactic emission from several processes contaminates the maps with an amplitude roughly proportional to the cosecant of the galactic latitude, compromising the observation of the CMB close to the galactic plane. In addition, a background of radio and infrared compact sources, galactic or extra-galactic,  contributes to the total emission even at high galactic latitudes.   Component separation consists in separating one or more of these sources of emissions  from the others in the data.

One of the main objectives of CMB experiments is the measurement of the CMB angular power spectrum $C_\ell$ which, with the assumption of statistical isotropy, can be estimated on a fraction of the sky. For this reason, many ground-based and balloon-borne experiments have concentrated their observations in ``clean'' regions of the sky, where galactic emission is low enough to impact negligibly the observations. For power spectrum estimation from full-sky observations, a safe approach consists in masking regions at low galactic latitude, and estimating power spectra on the cleanest regions of the sky. The impact of extragalactic point sources (evenly spread on the sky) on power spectrum estimates can be evaluated and corrected for using ancillary data (catalogues of known point sources and priors on the statistical distribution of sources).

Besides the power spectrum, the CMB map itself is interesting for several additional purposes:
\begin{itemize}
\item As a CMB template, to be subtracted from millimetre--wave observations when the scientific focus is on other emissions, or to be used for the calibration of other instruments;
\item To assess the statistical isotropy of the CMB and check the homogeneity and isotropy of the Universe on the largest scales;
\item To search for signatures of non-trivial topology, as that of a multi-connected universe \citep{2006MNRAS.369..240A,2007A&A...476..691C,2007PhRvL..99h1302N};
\item To search for correlations of the CMB map with other emissions \citep{2004ApJ...608...10N,2004MNRAS.350L..37F,2006MNRAS.372L..23C,2006MNRAS.369..645C,2006PhRvD..74d3524P,2007arXiv0704.0626M,2007MNRAS.377.1085R};
\item To search for signatures of non Gaussianity in the CMB \citep{2004ApJ...613...51M,2004ApJ...609...22V,2005MNRAS.362..826C,2006MNRAS.369..819C,2006MNRAS.369.1858M,2006MNRAS.365..265T,2006MNRAS.371L..50M,2007arXiv0706.2346W,2007arXiv0712.1148Y}.
\end{itemize}

Several CMB maps obtained with the WMAP data are available for such research projects.
The WMAP team has released part--sky foreground--reduced maps in the Q, V and W bands, and maps obtained by an Internal Linear Combination (ILC) of all WMAP channels \citep{2007ApJS..170..288H}. \citet{2003PhRvD..68l3523T} have produced CMB maps with WMAP one year data, and subsequently with three year data, based on an harmonic space ILC method.
\citet{2004ApJ...612..633E} have an alternate version of the ILC CMB map at 1 degree resolution on one year data.
\citet{2007arXiv0709.1037E} use a Gibbs sampler to draw 100 realisations of the  CMB under the posterior of  a model of CMB and foregrounds; their estimated CMB map is the average of these realisations for three year data. On three year data again, \citet{2007ApJ...660..959P} use an ILC technique on 400 different pixel ensembles, selected by the spectral index of the foreground emission as estimated by the WMAP team using a Maximum Entropy Method (MEM).
More authors have addressed component separation on WMAP three year data and produced versions of a ``clean'' CMB map \citep{2007MNRAS.374.1207M,2007arXiv0706.3567S,2007arXiv0707.0469B}. More recently, \citet{2008arXiv0803.1394K} have obtained a CMB map from WMAP five year data.

All available maps suffer from limitations, some of which result from specific choices in the way the CMB map is produced. Several of these maps, for instance, do not fully exploit the resolution of the original observations. Some focus on cleaning the CMB from foregrounds at high galactic latitude, and are significantly contaminated by foregrounds in the galactic plane. Some are not full sky CMB maps. Finally, not all of the available maps have well characterised noise and effective beam. All these limitations impact their usefulness  for accurate CMB science.

In this paper, we address the problem of making a CMB temperature map which has the following properties :
\begin{itemize}
\item being full sky;
\item being as close as possible to the true CMB (minimum variance of the error) everywhere on the sky, and on all scales;
\item having the best resolution possible;
\item having well-characterised beam and noise.
\end{itemize}

In the following, we first review these requirements and their impact on a CMB cleaning strategy (Section~\ref{sec:generalities}). We then review and compare available maps in Section~\ref{sec:existing}. In Section~\ref{sec:method}, we describe and explain our ILC needlet method. The approach is tested and validated on realistic simulated data sets (Section~\ref{sec:simulation}) before applying it to WMAP data (Section~\ref{sec:application}). We then compare our CMB maps to the other existing maps, discuss the results, and conclude. 

This paper considers only \emph{temperature} maps. 

\section{General considerations}
\label{sec:generalities}

\subsection{Requirements}

We start with a review of the requirements above, and on the implication on the method to be used.

\subsubsection{Full sky?}

WMAP data are full sky. We wish to devise a method which allows recovering an estimate of CMB emission everywhere, including in the galactic plane, and even (as much as possible) in the galactic centre as well as in pixels strongly contaminated by compact sources. 

The large scale correlation properties of the CMB make it possible to estimate the CMB emission even in unobserved regions, by some kind of interpolation. This is incidentally what is obtained with the Gibbs sampling technique of  \citet{2007arXiv0709.1037E}. Equivalent in spirit although quite different in implementation is the use of an ``inpainting'' method as that of \citet{inpainting}. Such interpolation methods alone are not fully satisfactory, as they discard information. In particular they do not allow recovering small scale CMB features in the mask. This is obvious, for instance, in the Gibbs-sampling average map  of  \citet{2007arXiv0709.1037E}.

At the opposite,  one may try  to separate components in the galactic plane independently of what is done at higher galactic latitudes, since levels and properties of foreground emission depend strongly on sky direction. \citet{2007ApJS..170..288H} and \citet{2003PhRvD..68l3523T} divide the sky into several independent regions, perform component separation independently in these regions, and then make a composite map by stitching together these independent solutions. Such approaches discard information (zone-to-zone correlations) and require careful treatment at the zone borders to avoid discontinuities and ringing.

A good method should perform well on both counts:  localised processing and  full exploitation of large scale correlations of the CMB and of galactic foregrounds. This can be achieved with a spherical wavelet or \emph{needlet} analysis of the maps (using the tools developed in \citet{2007arXiv0707.0844M} and \citet{2007arXiv0706.2598G}), which is our approach in the present work. 

\subsubsection{Minimum variance?}

Recovery of a CMB map can be conducted following various objectives,
quantified by different ``figures of merit''.  In this work, we
choose, as most authors, to minimise the variance of the difference
between true and recovered CMB (this is the \emph{overall} error; it
includes additive noise, foregrounds, and even multiplicative errors
affecting the CMB itself).

This choice alone does not fully characterise the method to be
used. The best theoretical solution also depends on the model of the
data. An overview of existing methods can be found in the review by
\citet{2007astro.ph..2198D}.

In this paper, contrarily to other approaches which rely heavily on the structure of the data as described by a model, for instance a noisy linear mixture of independent components \citep{2003MNRAS.346.1089D}, we assume as little as possible about the foregrounds and the noise. In fact, we assume nothing except the following:
\begin{itemize}
\item The WMAP data are well calibrated with known beam in each
 channel;
\item The instrumental noise in all WMAP maps is close to being
 Gaussian and uncorrelated; its pixel-dependent level is approximately known;
\item The CMB anisotropies emission law is known to be the derivative
 with respect to temperature of a T=2.725K black-body;
\item To first order, the template of CMB anisotropies is well
 represented by a Gaussian stationary random field, the spectrum of
 which is given by the WMAP best fit (as will be seen later--on, this last assumption is needed
 only to derive  the Wiener filter; it  is not necessary
 for our needlet ILC map).
\end{itemize}

These assumptions lead us to consider an ``Internal Linear Combination'' (ILC) method, followed by a Wiener filter to minimise the error integrated over all scales. 

\subsubsection{Best resolution?}

The WMAP data comes in five frequency channels with varying resolution. To make the best of the data, we need a method which uses the smallest scale information from the W band, and information from additional bands (V, then Q, then Ka, and finally K) at increasingly larger scales. Multi-scale tools are well suited for this purpose. 

The ``best possible resolution'' is not a well defined concept (and not necessarily the resolution of the W band). Indeed, there is a conflict between best possible resolution and minimum variance, as one can smooth or deconvolve a map arbitrarily in harmonic space, reducing or increasing the total noise variance in the process. Here, we make a map at the resolution of the W channel over the full sky, leaving open the option to filter this map if needed to reduce the noise -- or deconvolve it for better angular resolution. Note that additional global filtering or deconvolution does not change the signal to noise per mode (only the integrated S/N).

The minimum variance map is obtained with a Wiener filter, which
smoothes the map depending on the signal to noise ratio. As the noise
and the contaminants are inhomogeneous, there is a strong motivation
for the smoothing to depend on the location on the sky (the optimal
solution to the resolution--variance trade--off depending on the
contamination level, which is local). If we relax the constraint about
beam homogeneity, again, spherical needlets offer a natural way to
obtain such location--dependent smoothing.  
In the present, however, in order to preserve the constancy of the
effective beam over the sky, we implement the Wiener filter in harmonic space.

\subsubsection{Accurate characterisation?}

A fully accurate characterisation of the beam and noise is not
straightforward, in particular because of the limited knowledge about
the original frequency maps, which automatically propagates into the
final CMB map.  This work makes several approximations about beams and noise.
Beams are assumed symmetric and therefore described by the $b_\ell$
transfer functions provided by the WMAP team.
The instrumental noise is assumed uncorrelated, although non stationary.
Analytical analyses and Monte-Carlo simulations are used to
characterise the residual noise of the final map, as well as to estimate the
contribution of residual foregrounds, and biases if any. This will be detailed further later--on.

\subsubsection{Noise}

Throughout this paper, the term ``noise'' typically stands for all sources of additive error, i.e. instrumental noise {\emph{and}} foregrounds.

\subsection{Evaluation and comparison of reconstructed CMB temperature}

We briefly discuss here the tools used for characterising and comparing
CMB temperature maps.

\subsubsection{Map description}

A pixelised map is fully characterised by the specification of:
\begin{itemize}
\item A set of temperature values $y_p$ in a number of pixels (here indexed by $p$);
\item The effective beam at each pixel $p$, which in the most general case is a function $b_{p,p'}$;
\item The noise $n_p$, the statistical properties of which, in the Gaussian case, are fully described by a noise covariance matrix $N_{p,p'}$.
\end{itemize}
The map value $y_p$ is then linked to the true signal value $s_p$ by:
\begin{equation}
y_p = \sum_{p'}b_{p,p'} s_{p'} + n_p
\end{equation}

The full characterisation of a given CMB map requires the specification of the additive noise $n_p$ and of the response $b_{p,p'}$. When the beam is stationary over the sky and symmetric, which we assume in this work, it is fully specified by the coefficients $b_\ell$ of the expansion of the beam in Legendre polynomials.

\subsubsection{Assumptions}

Throughout this paper, the beam is assumed symmetric.  Although
this is an approximation, most pixels of the WMAP map are ``visited''
by any particular detector through a wide range of intersecting scans.
The average beam in that pixel then is an average of the physical beam
over many orientations, which makes the symmetry assumption
reasonable.

In addition, in absence of any specific localised processing, the beam is assumed to be invariant over the sky. 

With the above two assumptions, the effect of beam convolution is best represented in harmonic space, with a multiplicative coefficient $b_\ell$, independent of $m$, applied to the harmonic coefficients $a_{\ell m}$ of the map. We assume perfect beam knowledge as well as perfect calibration, so that no multiplicative uncertainty is attached to the map description (the beam integral, approximated as $\sum_{p'} b_{p,p'}$, is equal to unity independently of $p$, or, equivalently, the value of $b_\ell$ for $\ell = 0$ is assumed to be exactly unity).

The noise $n_p$ of the original WMAP maps, for each frequency channel and each differencing assembly, is assumed uncorrelated from pixel to pixel, i.e. $N_{p,p'} = \langle n_pn_{p'} \rangle = \delta_{pp'} \sigma^2_p$. The variance $\sigma^2_p$ is pixel dependent because of uneven sky coverage. Noise is non-stationary, but assumed to be Gaussian distributed.

\subsubsection{Comparison of maps at different resolution}

The comparison of CMB maps is meaningful only if the maps are at the same resolution. As long as beam transfer function does not vanish at any useful $\ell$ (which is always the case for Gaussian beams, and is also the case for WMAP best fit beams for all meaningful ranges of $\ell$), the resolution of any map can be changed to anything else by harmonic transform and multiplication by the ratio of the beam transfer functions. This property is widely used throughout this paper. 

\subsubsection{Masking}

We define ``tapered'' regions of the sky for map comparison at varying galactic latitude. In particular, we define a Low Galactic Latitude (LGL) region and a complementary High Galactic Latitude (HGL) region. The LGL region, used to evaluate results in the galactic plane, cuts completely all data above 30 degrees galactic latitude (and below $-$30 degrees), and has a 15 degree transition zone with a cosine square shape. All pixels at absolute galactic latitudes below 15 degrees are kept with a coefficient of 1. The HGL region is the complementary, i.e. HGL = 1$-$LGL. These ``tapered'' regions allow the computation of local power spectra with negligible spectral leakage of large scale power into small scales.

\subsubsection{Power spectra comparison}
\label{sub:powerspectra}

For a given beam (i.e. multiplicative response as a function of $\ell$), the comparison of the total map power as a function of $\ell$ (i.e. of the power spectra of the maps), is a direct figure of merit. The lower the power spectrum, the better the map.

Power spectra are computed independently for different regions of the sky (e.g. inside or outside the galactic plane).
To minimise aliasing due to sharp cuts, we use masks with smooth
transitions, defining the LGL and HGL regions described above.

The power spectrum of a map is evaluated as:
\begin{equation}
\widehat{C}_\ell = \frac{1}{(2 \ell + 1) \alpha} \sum_{m=-\ell}^{\ell} | a_{\ell m} |^2
\end{equation}
$\alpha$ is a normalising factor computed as the average value of  the squared masking coefficients.\footnote{The masking coefficient is simply 1 in regions kept for power spectrum computations, 0 in regions masked, and between 0 and 1 in the transitions.}

Power spectra estimated directly in this way from a masked sky are
unreliable for modes corresponding to angular sizes larger than the
typical size of the zone of the sky retained for computation.

\subsection{Methods}

We now give a brief introduction to the two main methods used in this paper (ILC and Wiener filtering). Many other methods exist for CMB cleaning (or component separation in general), which assume varying degrees of prior knowledge about sky emission, and model the data in different ways. These methods are not discussed nor used in this paper. For a review, see \citet{2007astro.ph..2198D}.

\subsubsection{The ILC}

The data are modelled as

\begin{equation}
\vec{x} = \vec{a} s +  \vec{n}
\end{equation}
where $\vec x$ is the vector of observations (e.g. five maps), $\vec a$ is the response to the CMB for all observations (e.g. a vector with 5 entries equal to 1 if WMAP data only are considered) and $\vec n$ is the noise. Here it is assumed that all observations are at the same resolution.

The ILC provides an estimator $\hat{s}_{\rm ILC}$ of $s$ as follows:
\begin{equation}
\hat{s}_{\rm ILC} = \frac{\vec{a}^t \, {\widehat{\tens{R}}}^{-1}}{\vec{a}^t \, {\widehat{\tens{R}}}^{-1} \, \vec{a}} \, \vec{x}
\label{eq:ILC}
\end{equation}
where ${\widehat{\tens{R}}}$ is the empirical covariance matrix of the observations (e.g., a $5 \times 5$ matrix when 5 channels are considered).

Note that the ILC solution of Equation~\ref{eq:ILC} is the linear filter which minimises the total variance of the output map, under the condition that the filter has unit response to the signal of interest (the signal with emission law given by vector $\vec{a}$).

The details of the method for its implementation in the context of this work are further discussed in Section~\ref{sec:method}.

\subsubsection{Wiener filtering}
\label{sub:wiener}

Given a single CMB map of known beam (assumed to be constant over the
sky), it is possible to minimise the contamination by noise and
foregrounds by (one-dimensional) Wiener filtering. The data is
modelled as $x=s+n$, where now $x$ is a single map, $s$ the true CMB
and $n$ the noise.  The Wiener filter gives to individual ``modes'' a
weight proportional to the fraction of signal power in that mode, i.e.
\begin{equation}
\hat{s}_{\ell m} = \frac{b_\ell^2 C_\ell}{b_\ell^2 C_\ell + N_\ell} \, x_{\ell m}
\label{eq:wiener-1d}
\end{equation}
where $b_\ell^2 C_\ell$ and $N_\ell$ are the power spectra of the (smoothed) CMB and of the noise (including smoothed foregrounds or foreground residuals) respectively, and $x_{\ell m}$ is the original noisy CMB map. It should be noted that if the CMB and the noise are uncorrelated, then $b_\ell^2 C_\ell + N_\ell = X_\ell$ is the power spectrum of the map $x_{\ell m}$, and the Wiener filter can be estimated directly using only a prior on the CMB power spectrum (assuming $C_\ell$ is known), and estimating $X_\ell$ on the map itself.

Wiener filtering in harmonic space minimises the variance of the error
in the map if signal and noise are Gaussian and stationary.
For non-stationary contaminants, the Wiener
filter~(\ref{eq:wiener-1d}) is still meaningful, but is no longer
optimal.  Some efficiency may be regained by an implementation in
another domain than the harmonic space (e.g. needlets), but this is not investigated further in this paper.

\section{Evaluation and comparison of available maps}
\label{sec:existing}

Before describing how to make yet another CMB map from WMAP data, we review the existing maps and evaluate in which respect they can be improved.

\begin{table*}[tb]
\caption{Available CMB maps.}             
\label{tab:allmaps}      
\centering          
\begin{tabular}{c c c l r }     
\hline\hline
NAME & resolution & data used & Reference & URL \\       
\hline                    
WILC1 & $1^\circ$ &  1-yr & \citet{2003ApJS..148...97B} & {http://lambda.gsfc.nasa.gov/product/map/dr1/imaps\_ILC.cfm} \\
TILC1 & W channel &  1-yr & \citet{2003PhRvD..68l3523T} & { http://space.mit.edu/home/tegmark/wmap/cleaned\_map.fits} \\
EILC1 & $1^\circ$ &  1-yr & \citet{2004ApJ...612..633E} & {http://www.astro.uio.no/$\sim$hke/cmbdata/WMAP\_ILC\_lagrange.fits} \\
\hline
WILC3 & $1^\circ$ &  3-yr & \citet{2007ApJS..170..288H} & { http://lambda.gsfc.nasa.gov/product/map/dr2/ilc\_map\_get.cfm} \\
EGS3 & $3^\circ$ &  3-yr & \citet{2007arXiv0709.1037E} & {http://www.astro.uio.no/$\sim$hke/gibbs\_data/cmb\_mean\_stddev\_WMAP3\_n64\_3deg.fits}\\
PILC3 & $1^\circ$ &  3-yr & \citet{2007ApJ...660..959P} & {http://newton.kias.re.kr/$\sim$parkc/CMB/SILC400/SILC400\_bc.fits}\\
TILC3 & W channel &  3-yr &  & {http://space.mit.edu/home/tegmark/wmap/cleaned3yr\_map.fits} \\
\hline
WILC5 & $1^\circ$ & 5-yr & \citet{2008arXiv0803.0715G} & { http://lambda.gsfc.nasa.gov/product/map/dr3/ilc\_map\_get.cfm} \\
KILC & $1^\circ$ & 5-yr & \citet{2008arXiv0803.1394K} & { http://www.nbi.dk/$\sim$jkim/hilc/} \\
\hline                  
\end{tabular}
\end{table*}

We start with a discussion of the existing methods, identifying for each of them strengths and 
weaknesses of the approach, and their foreseeable consequences. Available CMB maps obtained from WMAP data are compared in terms
of resolution, of the estimated contamination by foregrounds and of noise level. 
In absence of an absolute reference, discrepancies between available maps are also evaluated. 
This comparison permits to estimate typical uncertainties, and to outline
the ``difficult regions'' for CMB cleaning (which, unsurprisingly, are
mostly located close to the galactic plane).
We also look specifically for residuals of galactic contamination by comparing the power spectrum of the reconstructed CMB map at high and at low galactic latitudes. Significant discrepancies between the two are interpreted as indicative of a residue of foreground emission.

\begin{figure*}[htb]
 \centering
 \resizebox{0.49 \hsize}{!}{\includegraphics[angle=90]{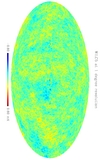}}
 \resizebox{0.49 \hsize}{!}{\includegraphics[angle=90]{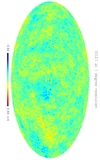}}
 \resizebox{0.49 \hsize}{!}{\includegraphics[angle=90]{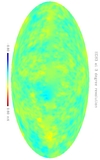}}
 \resizebox{0.49 \hsize}{!}{\includegraphics[angle=90]{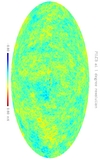}}
 \resizebox{0.49 \hsize}{!}{\includegraphics[angle=90]{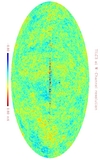}}
 \resizebox{0.49 \hsize}{!}{\includegraphics[angle=90]{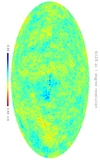}}
 \caption{This figure shows a selection of presently available CMB maps from WMAP data. Contamination by galactic emission is visible in all of them at various levels, except for the 3-degree resolution EGS3 map (for which a galactic cut was applied and then filled by a plausible CMB extrapolated from higher galactic latitude data).}
 \label{fig:cmbmaps}
\end{figure*}


\subsection{Available maps}

\subsubsection{The WMAP ILC}

The ILC maps obtained by the WMAP team (denoted as WILC1, WILC3 and WILC5 hereafter, depending on whether they are obtained using one year, three year or five year data) are described in \citet{2003ApJS..148...97B}, \citet{2007ApJS..170..288H}, and \citet{2008arXiv0803.0715G} respectively. 

For the three year and five year maps, the original frequency maps are smoothed to a common resolution of one degree. The sky is subdivided into 12 regions. One big region covers most of the sky at moderate to high galactic latitudes. The rest of the sky, concentrated around the galactic plane, is divided into 11 regions of varying galactic emission properties (amplitude and colour). An Internal Linear Combination is performed independently in each of these regions. A full sky composite map is obtained by co-adding the maps of the individual regions (with a $\simeq 1$ degree transition between the zones to avoid sharp edge effects). Finally, a bias (unavoidable consequence of empirical CMB-foreground correlation) is estimated by Monte-Carlo simulations, and subtracted from the composite map, to yield the final CMB map.

The three year and five year ILC maps differ from the one year map in several respects. The most significant is the recognition of the existence of a bias, and the attempt at correcting it using simulations. The limitations of the maps include their resolution (one degree), and the use of small regions in the ILC, which is bound to cause more bias than necessary on large scales (comparable to patch sizes), as well as edge effects. This results in discontinuities between regions, obvious for instance in the estimated bias map shown in \citet{2007ApJS..170..288H}. 

It is worth mentioning that in the method used by the WMAP team, the coefficients of the linear combination used over most of the sky (region 0, which corresponds to the largest part of the sky at high galactic latitudes and a few low galactic latitude patches, and region 1, in the galactic plane but away from the galactic center) are set using only a small subset of the data inside the Kp2 cut (where the galactic emission is the strongest). This choice favours the rejection of galactic contamination, at the price of sub-optimal weighting of the observations in regions where the error is dominated by noise. It also assumes that the emission laws and relative power of the different foregrounds are the same in these regions, which is a strong (and probably wrong) assumption.

Furthermore, the ILC weights are set by minimising the variance of the map at one degree resolution. Modes at higher $\ell$ get very sub-optimal weighting, as they do not contribute significantly to the total variance of the one degree map. The K and Ka bands, in particular, contribute respectively to about 0.156 and $-$0.086 for region 0 (the largest one) for WILC3. As a consequence, the final ILC map can not be meaningfully deconvolved to better resolution than about 1 degree (because this would blow-up small scale noise coming from the lowest frequency channels).  

Finally, there is also an unsatisfactory degree of arbitrariness in the choice of the regions, which depend on priors about  foreground emission, and are somewhat elongated across the galactic plane for no particular reason. Although none of these choices is unreasonable, the impact of this arbitrariness on the final map is difficult to evaluate.

For all these reasons, the WILC maps leaves considerable margin for improvement. We aim, in particular, at obtaining a CMB map with better angular resolution, and with a better handling of non-stationarity and scale dependence of the contamination (foregrounds and noise).

\subsubsection{The WMAP foreground-reduced maps}
\label{sub:forered}

For temperature power spectrum analysis, the WMAP team has used part-sky foreground-reduced maps. The processing for foreground removal for the three year and five year releases is described in  \citet{2007ApJS..170..288H}. Model templates for galactic emission are fitted to the Q, V and W WMAP channels outside of the Kp2 mask. A linear combination of synchrotron, free-free and dust, based on this fit, is then subtracted from the full sky Q, V, W maps.

In this procedure, a first galactic template, supposed to correspond to a linear combination of synchrotron and free-free emission, is obtained from the difference between the K and Ka bands. This template is produced at one degree resolution. An additional free-free template is obtained from H$\alpha$ emission \citep{2003ApJS..146..407F} corrected for dust extinction \citep{2003ApJS..148...97B}. A dust template is obtained from model 8 of \citet{1999ApJ...524..867F}. ``Clean'' Q, V and W maps are obtained by decorrelation of these templates from the original Q, V and W observations.

The main limitation of this approach is that the model used is insufficient to guarantee a good fit of the total foreground emission simultaneously inside and outside of the Kp2 mask. As a consequence, the maps produced are heavily contaminated by foregrounds in the galactic plane, the priority being given to higher galactic latitudes, with the objective of obtaining a part-sky high quality map on which high multipole CMB power spectra could be estimated reliably.

In addition, the maps are likely to depend significantly on the prior model assumed. Here, the WMAP team chooses dust model number 8 of \citet{1999ApJ...524..867F}, and also ignores the plausible existence of anomalous dust emission. The exact impact of these \emph{a priori} decisions is difficult to evaluate.

As an additional drawback, we note that the method generates correlated noise in the foreground-reduced maps, originating either from K and Ka channel noise or from a background of radio sources. Finally, on supra degree scales, the K and Ka bands, which are the most sensitive ones, are used only to subtract foregrounds, whereas in the cleanest regions of the sky they would be more usefully used to estimate the CMB emission.

For all these reasons, WMAP foreground-reduced maps are not good CMB maps according to the criteria listed in the introduction.

\subsubsection{The ILC by Eriksen et al.}

\citet{2004ApJ...612..633E} have obtained a CMB map at 1 degree resolution with another implementation of the ILC. The map, denoted here EILC1, uses only one year data. 

An interesting remark from Eriksen et al. is that the amount of residual dust is high in the ILC maps -- the method being able to subtract only about half of the dust present in the W band. At scales larger than 1 degree, this lack of performance is likely to be due to the part of dust emission uncorrelated to low frequency galactic foregrounds. On the smallest scales, the situation is worse, as the low frequency WMAP channels do not have the resolution to help remove small scale dust emission from W band observations. 

For this reason, in the present work, we improve on dust removal by using, as an additional measurement, the IRIS 100 micron dust template obtained from a combination of DIRBE and IRAS maps \citep{2005ApJS..157..302M}. As compared to the map of Eriksen et al., we also aim at better angular resolution -- and, obviously, better sensitivity, achieved by using five year data sets.

\subsubsection{The Gibbs-sampling map by Eriksen et al.}

Recently, \citet{2007arXiv0709.1037E} have produced a low resolution (3 degree) CMB map from WMAP three year data, using a Gibbs sampling technique to explore the likelihood of a parametric model of CMB and foreground emissions. A CMB map is obtained as the average of 100 sample CMB maps drawn each at random according to the posterior distribution of the model parameters given the observations.

The free parameters in the model are spherical harmonic coefficients $a_{\ell m}$ of the CMB map, the CMB harmonic power spectrum $C_\ell$, monopole and dipole amplitudes in each WMAP band, the amplitude $a(\nu)$ of a dust template in each band, and amplitudes $f(p)$ and spectral indices $\beta(p)$ of a low-frequency foreground component, for each map pixel $p$.

The model is constrained by fixing the dust template at 94 GHz according to \citet{1999ApJ...524..867F}, by a prior on the low-frequency foreground spectral index, assumed to be close to that of synchrotron ($-3 \pm 0.3$), and by the constraint that the monopole and dipole coefficients are orthogonal to the (noise-weighted) pixel--averaged foreground spectrum.

In spite of a good fit of the assumed model to the data at high galactic latitudes, there are some strong limitations to the resulting CMB map, and hence to its usage:
\begin{itemize}
\item The result of the sampling is obtained assuming a parametric model of foreground emission. There is no possible way of validating the systematic errors due to mismodelling, except marginalising over all possible model skies. This would require a Monte-Carlo simulation which takes into account all uncertainties in the modelling, not only values of the parameters for a given parametric model, but also the choice of the parameter set to be used to model the foregrounds (varying the dust template according to uncertainties, assuming a different foreground model, etc...). This is not presently available;
\item The resulting map is at 3 degree resolution and HEALPix nside = 64, considerably worse than WMAP can do;
\item The data sets are cut with the Kp2 mask. Although a CMB is recovered in the mask by the average of the sample maps, the effective resolution inside the mask is lower than in the rest of the sky. In some sense, the Gibbs sampling technique (as implemented here) allows to recover in the mask what is predictable from the outside map (assuming stationarity of the CMB anisotropy field). It allows only for a limited  prediction of the CMB signal in the masked zone.
\end{itemize}

For all of these reasons, the Gibbs-sampling map of Eriksen et al. (hereafter EGS3) is not a good ``best CMB'' map according to our criteria.

\subsubsection{The ILC by Park et al.}

\citet{2007ApJ...660..959P} provide their own version (hereafter the PILC3 map) of a one degree resolution CMB map obtained by an ILC on WMAP three year data. The originality of their approach lies in the fact that they cut the sky into 400 pixel ensembles, selected from a prior on their spectral properties. The 400 ensembles are defined from $20 \times 20$ spectral index bins (20 for K-V spectral index, and 20 for V-W). This approach is motivated by the fact that ILC weights are expected to vary with varying foreground properties.

There is a weak point to this approach. The authors use, to define their pixel `bins', the MEM solution derived by the WMAP team. If the MEM solution is wrong for a given pixel, that pixel will automatically be classified in the wrong pixel ensemble, and be weighted using the weights of the wrong population of pixels. To some extent then, this binning forces the result of the ILC to match the prior assumptions given by the MEM results. In turn, the MEM solution uses as a prior the result of the WMAP ILC, which is subtracted from the WMAP observations prior to using the MEM method to separate galactic foregrounds.

As a consequence, the connection of the CMB map of Park et al. to the original WMAP data is far from direct. The map is bound to bear the signature of any arbitrary choice made before, in particular the choice of WMAP ILC regions, the 3-component model for galactic emission, and the MEM priors. For instance, discontinuities at the boundaries between the 12 regions of the WMAP ILC are clearly visible in the map of K-V spectral index used by the authors, as well as their group index (see Figures~3a and~4a of their paper).

Park et al. then investigate the error in their reconstructed map by Monte-Carlo simulations. However, they use as an input galactic emission template the very model obtained by the MEM. This means that in the simulations, the spectral index maps are ``exact''. Therefore, the simulations investigate accurately the errors only if the MEM solution is correct, which is not likely to be the case --at least not to the level of precision required for producing a CMB map useful for precision cosmology.

Our method described in Section~\ref{sec:method} uses as little prior information as possible, and aims at better angular resolution than 1 degree.

\subsubsection{The ``clean'' map of Tegmark et al.}

The approach of \citet{2003PhRvD..68l3523T}, on both one year and three year data, is the only work to date which aims at producing a CMB map with both full sky coverage and best possible resolution.

\citet{2003PhRvD..68l3523T} have performed a foreground analysis of
the WMAP one year maps, producing two high resolution maps of the
CMB. The first one, the ``clean'' map (hereafter TILC1), is obtained by
a variant of the ILC in which weights are allowed to vary as a
function of $\ell$, and are computed independently in nine independent
regions. The second, the ``Wiener'' map, is a Wiener-filtered version
of the same map, in which the Wiener filter is applied in harmonic
space, but independently in each zone.

This work by Tegmark et al. is an early attempt at finding linear
combinations of the WMAP data which vary both in harmonic space and in
pixel space.  The pixel variation of the weights is made using zones
which are defined according to the level of contamination by
foregrounds, as computed from WMAP map differences W-V, V-Q, Q-K and
K-Ka.
The authors do not specify exactly how the frequency range is divided
into $\ell$-bands.  Whereas the text seems to indicate that weights
are computed independently for each $\ell$, figures hint that the
weights are actually band-averaged, in 50 logarithmic bands
subdividing the multipole range. In the end, the authors obtain a CMB
``clean'' map with a ``beam corresponding to the highest-resolution
map band'', i.e. the beam of the W band.

The original paper describes the work done on the one year WMAP data. However maps for three year data are available on Max Tegmark's web site (see Table~\ref{tab:allmaps}). We use the three year map (TILC3) for comparison with our own solution. 

Although the approach of Tegmark and collaborators is quite good at
high galactic latitudes, we can see on Figure~\ref{fig:cmbmaps} that
it performs poorly in the galactic plane.  Also, the authors have not
removed detected point sources from the WMAP data before making the
ILC. As a result, their CMB map contains obvious point source
residuals, for instance around galactic longitude 305$^\circ$ and
latitude 57$^\circ$, where a 5 mK peak can be seen.

Our method, although bearing some similarity with that of Tegmark and
collaborators, aims at improving significantly the error
characterisation, as well as the quality of foreground cleaning in the
galactic plane.

\subsubsection{The ``Wiener'' map of Tegmark et al.}

In addition to their TILC map, \citet{2003PhRvD..68l3523T} publish a Wiener map (hereafter TW map), obtained from the TILC map by independent Wiener filtering in the 9 regions. This results in reduced integrated error in all regions, at the price of pixel-dependent extra smoothing.
The consequence of this filtering is an effective zone-dependent beam.

Because of this extra smoothing, it is difficult to compare the TW map with other maps. The most meaningful figure of merit for the Wiener map, anyway, is the actual power of the error (output map minus true CMB). This is unavailable for any useful up-to-date real data set. Additional discussion about Wiener-filtered maps is deferred to Section~\ref{sec:application}.

\subsubsection{The ILC map of Kim et al.}

More recently, \citet{2008arXiv0803.1394K} have made a CMB map from
WMAP five year data, using an ``harmonic'' ILC method (KILC5 hereafter).
Their method perform an ILC in the pixel domain but with
pixel-dependent weights. 
The ILC weights are not constant over predefined zones on the sky but
are computed as smooth weight maps defined in terms of an harmonic
decomposition (hence the qualification of the method).
More specifically, the weight maps are determined by minimising the
total output CMB map variance with the constraint that these maps have
no multipoles higher than $\ell_\mathrm{cutoff}$.  For stability
reasons, the KILC5 map is obtained with $\ell_\mathrm{cutoff}=7$.
Prior to computing ILC weights, all maps are deconvolved from their
beam (effectively blowing up noise on small scales, in particular for
the lowest frequency channels). Then, the channels are combined using
map modes for $\ell < 300$.

With the above choices, the reconstructed map can not be good on small
scales.  As the authors notice themselves, using small scale modes
results in minimisation of noise rather than foregrounds (and,
obviously, rejecting the low-frequency observations, which are the
noisiest on small scales after deconvolution from the beam).  Better
results could probably be obtained by estimating weight maps for
different bands of $\ell$.  In essence, this is what our needlet ILC
method permits to achieve.

As a last comment, we note that limiting the number of modes of the
weight maps to $\ell \leq 7$, reportedly for reasons of singularity of
the system to be solved, results in spatial coherence of the weights
on scales of about 35 degrees.  The galactic ridge, however, is about
1~degree thick.  Hence, the spatial variability of the ILC weights
achieved by \citet{2008arXiv0803.1394K} is not quite adapted to the
actual scale of foreground variation.  The needlet ILC method
presented in our paper, as will be seen later on, solves this issue in
a very natural way.

\subsubsection{Other maps}

Other authors have performed various foreground cleaning in the WMAP observations, producing a number of CMB maps for several different models of the foreground emission. \citet{2007arXiv0707.0469B} perform component separation on WMAP data using the CCA method described in \citet{2006MNRAS.373..271B}. \citet{2007MNRAS.374.1207M} obtain also several CMB maps, using the FastICA method \citep{2002MNRAS.334...53M}. None of the CMB maps obtained is full sky, nor publicly available yet. They are not considered further in this analysis. 

Finally, some foreground cleaning has also been performed by \citet{2007arXiv0706.3567S}. Their paper also includes an interesting analysis of the ILC bias. The primary goal of that work, however, is to compute the CMB power spectrum, rather than producing a CMB map. 

\subsubsection{Existing map summary}

Figure~\ref{fig:cmbmaps} shows six available maps, all displayed in
the same colour scale.  It illustrates the resolution and foreground
contamination of the various maps.  Table \ref{tab:allmaps} summarises
the main properties of the maps. Only the TILC1 and TILC3 maps are high
resolution attempts at component separation everywhere, including the
galactic plane, combining all WMAP observations. All other maps (WILC maps, EILC1, EGS3, PILC3, KILC5) are
at reduced resolution.

\subsection{Map comparison}

CMB maps produced from WMAP one year data have been compared by
\citet{2004ApJ...612..633E}, showing quite large differences, ranging from $-100$ to $100$~$\mu$K.
Similarly, \citet{2007ApJ...660..959P} compare their PILC3 map with the WILC3 and TILC3 maps at 1.4 degree resolution,
showing differences in excess of 40~$\mu$K. 
In the following, discrepancies between these various solutions are
further investigated.

\begin{figure}[htb]
 \centering
 \resizebox{\hsize}{!}{\includegraphics[angle=90]{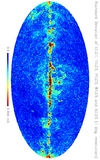}}
 \resizebox{\hsize}{!}{\includegraphics[angle=90]{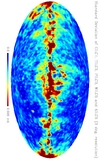}}
 \caption{Top: the standard deviation, per pixel for nside=512, of the EILC1, TILC3, PILC3, WILC5 and KILC5 maps, at 1$^\circ$ resolution. Bottom: same at 3$^\circ$ resolution. Note the different color scales for the two panels.}
 \label{fig:stddev-map}
\end{figure}

As a first step, we evaluate by how much the various maps at one degree resolution disagree on the full sky. The top panel of Figure~\ref{fig:stddev-map} shows the pixel based standard deviation of five such maps. In this evaluation, we exclude the EGS3 map (which is at 3 degree resolution, and is the source of additional variance in the Kp2 sky mask, where the CMB is estimated at even lower resolution).

All maps are obtained from observation of the same sky (sometimes starting from the exact same data set) and are smoothed to
the same resolution (one degree) where noise is small. Discrepancies originate essentially from
systematic differences in the methods. In the central
regions of the galactic plane, discrepancies are significantly higher
than 50~$\mu$K (with bright spots above 90~$\mu$K). In other sky
regions, they are typically in the 10-20~$\mu$K range, except on
compact spots again, where they are above 50~$\mu$K. The latter are
probably due to residuals of the emission from strong compact
sources. The discrepancy between the maps is larger in the ecliptic
plane, which is a signature of the impact of the instrumental noise, uneven because
of the WMAP scanning strategy. This is due to the different attention
given to minimising the contribution of instrumental noise on small scales (rather than foregrounds) in the final map (and to a much lesser extent to the difference of noise level between the various WMAP releases).

The bottom panel of Figure~\ref{fig:stddev-map} focuses the comparison to larger scales (3 degree beam). At high galactic latitudes, the discrepancies are below 10~$\mu$K except for a few localised regions (LMC, Ophiuchus complex) where they reach about 30~$\mu$K. Differences in the North Polar Spur, at the level of 10-15~$\mu$K, are also clearly visible. Close to and inside the galactic plane, systematic discrepancies significantly exceed 30~$\mu$K.

Including the EGS3 map for comparisons at high galactic latitudes does not change these conclusions. Pairwise comparisons of the available maps typically show the same level of discrepancies, which indicates that the variance of the solutions is not due to one single map in strong disagreement with all the others.

The conclusion of this comparison is that foreground residuals exist in the published CMB maps at the level of about 50 to 100~$\mu$K in the galactic ridge, 20 to 50~$\mu$K at low galactic latitudes, and 10 to 20~$\mu$K at higher galactic latitudes (above 30 degrees).

This observation calls for localised weightings, adapted to local properties of the foregrounds and the noise. This, however, is not easily compatible with the recovery of the largest modes of the CMB map, as pointed out before by \citet{2004ApJ...612..633E}, and as demonstrated by the discontinuities observed between the CMB solutions in the different regions when the sky is cut, as in the TILC, EILC1, and WILC maps.

Our approach, then, will be to vary the relative weightings on small scale for small scale CMB reconstruction, and keep the weighting uniform over large regions of the sky for the recovery of the largest scales. This can be achieved quite straightforwardly by  using the spherical needlets discussed in Section~\ref{sec:method}.

\section{The ILC needlet method}
\label{sec:method}

\subsection{The choice of the ILC}

It is striking that all the presently available full sky CMB maps derived from an analysis of the WMAP data have been obtained by an implementation of the ILC method.

The ILC, indeed, has many advantages:
\begin{itemize}
\item The method relies only on two very safe assumptions: the CMB emission law, and the fact that the CMB template is not correlated to foreground emission;\footnote{
In reality, it is likely that the CMB map actually \emph{is} somewhat correlated to the foregrounds (extragalactic point sources and SZ effect), because of the ISW effect. The implication of this is not studied further in the present paper.}
\item Under these assumptions, the method minimises the empirical variance of the reconstruction error;
\item The ILC is very simple in implementation.
\end{itemize}

\noindent
The ILC has also two major drawbacks.
\begin{itemize}
\item As noted for instance by \citet{2007ApJS..170..288H},  \citet{2007astro.ph..2198D}, \citet{2007arXiv0706.3567S}, empirical correlations between the CMB and the source of contamination results in a bias; this bias is discussed in more detail below;
\item In absence of a model of the contaminants (foregrounds and noise), it is not possible to predict the reconstruction errors, which somewhat annihilates the benefit of making very safe assumptions about the properties of the data set.
\end{itemize}

\subsection{The ILC bias}
\label{sec:ILC-bias}

The existence of a ``bias'' in maps obtained by an ILC method is a well established fact. The derivation of this bias (which is, in fact, the systematic cancelling of a fraction of the projection of the CMB map onto the vector space spanned by the noise realisations for all the considered input maps), is given in appendix.

The order of magnitude to keep in mind is that about $(m-1)$ ``modes'' of the original CMB, out of $N_p$, are cancelled by the ILC, where $m$ is the number of channels used, and $N_p$ the number of independent pixels or modes in the regions for which the ILC is implemented independently. Note however that when the signal is strongly correlated between pixels, the bias can be significantly larger --see the appendix for details.

The practical consequences are:
\begin{itemize}
\item a loss of CMB power, which has to be taken into account for power spectrum estimation;
\item an anti-correlation of the map reconstruction error with the real CMB sky.
\end{itemize}

The level of the bias induced by our method is investigated both theoretically (Appendix~\ref{app:bias}), and through Monte-Carlo simulations.

\subsection{Needlets}
\label{sec:needlets}

A \emph{frame} is a collection of functions with properties close to those of a basis. \emph{Tight frames} share many properties with orthonormal bases, but are redundant (see \cite{daubechies:1992} for details).

Needlets were introduced by \citet{narcowich:petrushev:ward:2006} as a particular construction of a wavelet frame on the sphere. They have been studied in a statistical context (e.g. \citet{baldi:etal:2006a,baldi_etal:2007}) and have also been used recently for cosmological data analysis problems (e.g. \citealt{2006PhRvD..74d3524P}).  The most distinctive property of the needlets is their simultaneous perfect localization in the spherical harmonic domain (actually they are spherical polynomials) and potentially excellent localization in the spatial domain.

We recall here the definition and practical implementation of the needlet coefficents, following the generalised formulation by \cite{2007arXiv0706.2598G}. Let $h_\ell^{(j)}, j \in \mcJ$ be a collection of window functions in the multipole domain, indexed by $j$. Suppose that for each scale $j$, $\xi^{(j)}_k$ is a grid of points (indexed by $k \in K^{(j)}$)
which satisfies an exact\footnote{or almost exact, for all practical purposes} quadrature condition with weights $\lambda_k^{(j)}$.  The needlets are axisymmetric functions defined by
\begin{equation}
 \label{eq:defneedlet}
 \psi^{(j)}_k(\xi)=\sqrt{\lambda^{(j)}_{k}}\sum_{\ell=0}^{\ell_{\max}}
 h^{(j)}_\ell L_\ell (\xi \cdot \xi^{(j)}_k ),
\end{equation}
where $L_\ell$ denote the Legendre polynomials.

Any square integrable function $f$ on the sphere can be analysed by the scalar products $\beta^{(j)}_k := \langle f , \psi^{(j)}_k \rangle$ of the function $f$ with \emph{analysis needlets}. All the needlet coefficients of scale $j$ are advantageously computed in the spherical harmonic domain, as the evaluation at points $\xi_k^{(j)}$ of a function whose multipole moment are simply $h^{(j)}_\ell a_{\ell m}$. These needlet coefficients, denoted $\gamma^{(j)}_k$, are given by:
$$
\gamma^{(j)}_k = \sqrt{\lambda^{(j)}_k} \, \beta^{(j)}_k
$$

Each field of needlet coefficients can in turn be convolved with some \emph{synthesis needlets}
\begin{equation}
 \label{eq:defneedlet_recons}
 \tilde\psi^{(j)}_k(\xi)=\sqrt{\lambda^{(j)}_{k}}\sum_{\ell=0}^{\ell_{\max}}
 \tilde h^{(j)}_\ell L_\ell (\xi \cdot \xi^{(j)}_k ),
\end{equation}
using the exact same procedure and leading to the map $X^{(j)}$ whose multipole moments are $h_\ell^{(j)}\tilde h_\ell^{(j)}a_{\ell m}$.  The analysis and synthesis operations are summed up as:\\

\noindent \emph{Analysis:}
$$
\begin{array}{ccccccc}
X &\stackrel{\textrm{SHT}}{\longrightarrow} & a_\lm &
\stackrel{\times}{\longrightarrow}& h^{(j)}_\ell a_\lm &
\stackrel{\textrm{SHT}^{-1}}{\Longrightarrow} & \gamma^{(j)}_k
\end{array}
$$
\noindent \emph{Synthesis:}
$$
\begin{array}{ccccccc}
\gamma^{(j)}_k & \stackrel{\textrm{SHT}^{-1}}{\Longrightarrow} & 
h^{(j)}_\ell a_\lm & \stackrel{\times}{\longrightarrow} & \tilde h^{(j)}_\ell h^{(j)}_\ell a_\lm & \stackrel{\textrm{SHT}}{\Longrightarrow} & X^{(j)}
\end{array}
$$
Double arrows denote as many transforms as scales in $\mcJ$.
If $X$ is band-limited to $\ell \leq \ell_{\max}$ and if the
reconstruction condition $ \sum_j h_\ell^{(j)}\tilde h_\ell^{(j)} = 1$
holds for all $\ell \leq \ell_{\max}$, then the complete process
yields a decomposition of $X$ in smooth maps, namely
\begin{equation}
 \forall \xi , \;\;\;\; X(\xi) = \sum_j X^{(j)}(\xi) \ 
\end{equation}

Note that the existence of a fast inverse spherical harmonic transform
using the quadrature points $\xi_k^{(j)}$ is required in practice, and
that HEALPix pixels and weights fulfil the quadrature condition only
approximately.  Further details can be found
in~\cite{2007arXiv0706.2598G}, with an extensive discussion on the
choice of the spectral window functions.

A key feature of the needlet decomposition follows from the
localization of the analysis functions which allows for localised
processing (such as denoising, signal enhancement, masking etc.) in
the needlet coefficient domain, \emph{i.e.}  applying some non-uniform
transforms to the coefficients $\gamma_k^{(j)}$.


\subsection{The method}

The method implemented in this work, and applied both to simulations and to the real WMAP data sets (for all releases), consists in the following steps:
\begin{itemize}
\item We start with the data set consisting of band-averaged temperature maps from WMAP (simulated or real data), to which we add the IRIS 100 micron map;
\item WMAP-detected point sources are subtracted from the WMAP maps; 
\item We apply a preprocessing mask, in which a very small number of very bright, compact regions, are blanked (see Table~\ref{tab:sourcelist}); blanked regions are filled-in by interpolation; this is done only on the real WMAP data.
\item All maps are deconvolved to the same resolution (that of the W channel\footnote{A noise weighted average beam is obtained from the W1 W2 W3 W4 beam coefficients provided by the WMAP team}); this operation is performed in harmonic space;
\item Maps are analysed into a set of needlet coefficients $\gamma^{(j)}_k$ following the method described in \ref{sec:needlets};
\item For each scale, the covariance matrix $\widehat{R}$ of the observations is computed locally (using an average of $32 \times 32$ needlet coefficients);
\item The ILC solution is implemented for each scale in local patches;
\item An output CMB map is reconstructed from the ILC filtered needlet coefficients; this map constitutes our main CMB product;
\item That map is Wiener-filtered in harmonic space, to make an alternate CMB map with lower integrated error (our best guess CMB map).
\item In parallel, the actual ILC filter used on the analysed data set is applied to 100 different simulations of the WMAP noise, to estimate the noise contribution to the final map.
\item The level of the biasing, which depends on the geometry and not much on the actual templates of CMB and foreground and noise, is estimated on a set of fully simulated data.
\end{itemize}
Each of these steps is described in more detail below.

\subsection{Point source subtraction}
Strong point sources in the input data set typically leave detectable residuals in the output ILC map, and twist at the same time the estimation of the background, thus conducting to lower rejection of other contaminants. On the other hand, their specific shape usually allows effective estimation and removal by other methods. For the purpose of this study we used information from the WMAP source catalogue (\citet{2007ApJS..170..288H}) which provides characterisation for all point sources detected above a \(5\sigma\) threshold away from the galactic plane. 

For all sources identified, we subtract from the input maps a Gaussian profile at the given position and with the given flux. Conversion factors between flux density and Gaussian amplitude, as well as the FWHM of the Gaussian profile are taken from Table 5 of \citet{2003ApJS..148...39P} for one year data, and corresponding updates for the more recent releases.

For the simulated data set, the subtraction of detected point sources is mimicked by removing from the simulation all sources above 1~Jy (independently in all channels).

\subsection{Blanking of compact regions}
\label{sub:blank}

In addition to the point sources subtracted above, some compact regions of strong emission (mostly in the galactic plane) exceed the rejection capabilities of the method used in this analysis, because they are too local and/or too specific. Their contribution in the wings of the needlets also contaminate the solution far from the centre of the sources. Those sources cannot be satisfactorily subtracted in the same way than the previous ones, either because they are not strictly speaking point-like, or because they are bright enough that small departures of actual beam shapes from the Gaussian model used in the subtraction step leave significant residuals.
As they represent only a very tiny fraction of the sky (we single out eleven such sources), we blank out these regions in all WMAP channels, cutting out circular patches adapted to the size of the beam and of the source.
Table \ref{tab:sourcelist} gives the list of those regions with their main characteristics.

To reduce local pollution of the needlet coefficients by the sharp cut, the small blanked regions are filled in by a smooth interpolation, so that fluctuations at a larger scale than the hole size are at least coarsely reconstructed. More precisely, interpolation is made by diffusion of the boundary values inside the hole. 

Although this masking and interpolation has no reason to be optimal, it is an efficient way of reducing the impact of very strong sources on their environment. The CMB inside the masked patches is recovered (to some extent) both by the interpolation of original maps (which avoids sharp discontinuities) and by the needlet decomposition and ILC reconstruction. The masked region is tiny: 0.058\% of the sky in the K channel (the most affected).

\begin{table}
 \centering
 \begin{tabular}{|c|c|c|}
   \hline
   Name & Galactic coordinates & Type \\
   \hline
   Crab neb & 184.5575 -05.7843 & SNR\\
   sgr A & 000.064 +00.147 & Radio-Source \\
   sgr B & 000.599 +00.002 & Radio-Source \\
   sgr C & 359.4288 -00.0898 & HII region \\
   sgr D & 001.131 -00.106 & Molecular cloud \\
   Orion A & 209.0137 -19.3816 & HII region \\
   Orion B & 206.5345 -16.3539 & Molecular cloud \\
   Omega neb & 015.051 -00.674 & HII region \\
   Cen A & 309.5159 +19.4173 & QSO \\
   Cas A & 111.735 -02.130 & SNR \\
   Carina neb & 287.6099 -00.8542 & HII region \\
   \hline
 \end{tabular}
 \caption{List of compact regions blanked in the pre-processing step. A circular patch centered on the source, of radius 75, 55, 45, 45, 34 and 35 arcminute for the K, Ka, Q, V, W and IRIS 100 \(\mu m\) bands respectively, is masked. The masked regions are then filled with an extrapolation of edge values.}
 \label{tab:sourcelist}
\end{table}

\subsection{Needlet decomposition}

The original observations (WMAP and IRIS) are decomposed into a set of filtered maps represented by their spherical harmonic coefficients:
\begin{equation}
a_{\ell m}^{(j)} = h_{\ell}^{(j)} a_{\ell m}
\end{equation}
where $a_{\ell m}$ are the spherical harmonics coefficients of the original map, and $a_{\ell m}^{(j)}$ those of the same map filtered by the window function $j$.
Needlet coefficients $\gamma^{(j)}_k$ are obtained as the value of the filtered map at points $\xi_k$.

\begin{figure}[htb]
 \centering
 \resizebox{\hsize}{!}{\includegraphics[angle=0]{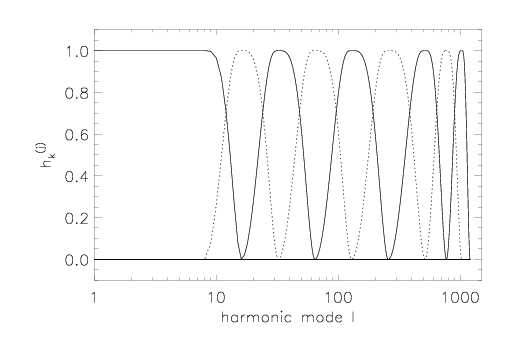}}
 \caption{The spectral bands used in this work for the definition of the needlets.}
 \label{fig:bands}
\end{figure}

For each scale $j$, the coefficients $\gamma^{(j)}_k$ are computed on a HEALPix grid at some value of {\tt nside}, compatible with the maximum value of $\ell$ of band $j$. We use for {\tt nside}$(j)$ the smallest power of 2 larger than $l_{\rm max}/2$, with a maximum of 512. Details about the bands used are given in table \ref{tab:needlets} and figure \ref{fig:bands}.

\begin{table}
 \centering
 \begin{tabular}{|c|c|c|c|}
   \hline
   Band $(j)$ & $\ell_{\rm min}$ & $\ell_{\rm max}$ & {\tt nside}$(j)$  \\
   \hline
   1 & 0 & 15 & 8 \\
   2 & 9 & 31 & 16 \\
   3 & 17 & 63 & 32 \\
   4 & 33 & 127 & 64 \\
   5 & 65 & 255 & 128 \\
   6 & 129 & 511 & 256 \\
   7 & 257 & 767 & 512 \\
   8 & 513 & 1023 & 512 \\
   9 & 769 & 1199 & 512 \\
   \hline
 \end{tabular}
 \caption{Spectral bands used for the needlet decomposition in this analysis. Needlet coefficient maps are made at different values of {\tt nside}, given in the last column.}
 \label{tab:needlets}
\end{table}

\subsection{ILC implementation on needlet coefficients}
\label{sec:ilc-impl-needl}

The general idea is to implement independently the ILC on subsets of the needlet coefficients $\gamma^{(j)}_k$. For a given scale, these coefficients come in the format of a set of HEALPix maps (one per frequency channel). The ILC is implemented locally in space and locally in $\ell$ as follows.

Covariance matrices ${\tens{R}}^{(j)}_k = \langle {\vec \gamma}^{(j)}_k {{\vec  \gamma}^{(j)}_k}^T \rangle $ for scale $j$ at pixel $k$ are estimated as the average of the product of the computed needlet coefficients over some space domain $\mathcal{D}_k$. In practice, for the present analysis, we make use of the hierarchical property of the HEALPix pixellisation, and compute covariance matrices as the average in larger pixels, corresponding to a HEALPix pixellisation with {\tt nside} $=$ {\tt nside}$(j)/32$. This provides a computation of the statistics by averaging $32^2 = 1024$ samples, which results in a precision of order 3 per cent for all entries of the ${\tens{R}}^{(j)}_k$ matrix. It implies an ILC bias of order $5/1024$ (for $m=6$ channels and $N_p=1024$ coefficients per domain on which the ILC filter is estimated independently (see appendix for details).

We denote as ${\widehat{\tens{R}}^{(j)} _{k \in \mathcal{D}_k} }$ the estimate of ${\tens{R}}^{(j)}_k$ obtained by averaging the value of ${\vec \gamma}^{(j)}_k {{\vec  \gamma}^{(j)}_k}^T$ in domain $\mathcal{D}_k$.

On the largest scales ($\ell \le 50$), the typical angular extent of a needlet is larger than 5 degrees, and the value of {\tt nside} for the map of needlet coefficients is less than 32. Covariance matrices are then computed on the full sky rather than on the largest possible HEALPix grid, i.e. $\mathcal{D}_k$ is the complete sky, rather than one of the 12 basis Healpix pixels.

Using these covariance matrices, the ILC is implemented using equation \ref{eq:ILC} for each domain. The estimated CMB needlet coefficients are:
\begin{equation}
\left [  \widehat{\gamma}^{(j)}_k \right ] _{\rm CMB} = 
\frac	{  \vec{a}^T   \left [                {\widehat{\tens{R}} ^{(j)} _{k \in \mathcal{D}_k} }              \right ] ^{-1}      } 
	{  \vec{a}^T   \left [                {\widehat{\tens{R}} ^{(j)} _{k \in \mathcal{D}_k} }              \right ] ^{-1}     \vec{a}  } 
	\, \vec{\gamma}^{(j)}_k 
\end{equation}

\subsection{Full map reconstruction}

The full CMB map reconstructed from this set of needlet coefficients is our basic needlet ILC (NILC) CMB map. 

\subsection{Final Wiener filtering}
\label{sec:post-wiener}

For a number of purposes, in particular subtraction of an estimate of the CMB to study other emissions, it is interesting to use, instead of our ILC map at the resolution of the WMAP W channel, a map with minimal error. Such a map is obtained from the ILC map by  one-dimensional Wiener filtering.

As a last processing step towards a minimum variance CMB map, we thus Wiener-filter our CMB map, to get rid of the large noise contamination at high $\ell$. The Wiener filter is performed in harmonic space as described in \ref{sub:wiener}.

The harmonic Wiener filter is given by formula \ref{eq:wiener-1d}, i.e.  $w_\ell = b_\ell^2 C_\ell/(b_\ell^2 C_\ell+N_\ell)$. For its implementation, we need to know the relative power of CMB and noise. We assume that the best fit CMB power spectrum of the WMAP team is correct, hence $C_\ell$ is known. The beam factor $b_\ell$ is assumed perfectly known as well. The denominator $b_\ell^2 C_\ell + N_\ell$ can be estimated directly as the power spectrum of our output needlet ILC map.

In practice, we smooth the power spectra with $\delta \ell / \ell = 0.1$ to lower the variance of the power spectrum estimator on the output needlet ILC map. Even with this, the filter is poorly estimated for low modes, because of the large cosmic variance. As can be seen on Figure~\ref{fig:cmb-simu-spec}, the signal to noise ratio of our reconstructed map is expected to be quite high at low $\ell$. Therefore, the Wiener filter for low modes is expected to be very close to 1. For this reason, we set $w_\ell = 1$ for $\ell<200$, and use a linear interpolation between $\ell=200$ and $\ell=250$. 

\subsection{Noise level estimate}

The level of noise contamination (variance per pixel, and average power spectrum) in the output map is estimated by Monte-Carlo, using the average of 100 realisations of the WMAP noise maps. For each initial set $(i)$ of five noise maps (one noise map per WMAP channel), a single output noise map $n_p^{(i)}$ is obtained by performing the needlet decomposition of the initial noise maps, and filtering needlet coefficient maps with the same filter as that used on the single full simulated data set.

Denoting as $n_p^{(i)}$ and $n_{\ell m}^{(i)}$ respectively the pixel value and the harmonic space value of the noise map number $i$, we compute:

\begin{equation}
\sigma^2_p = \frac{1}{N_i} \sum_i \left ( n_p^{(i)} \right )^2
\end{equation}
and
\begin{equation}
\sigma^2_\ell = \frac{1}{N_i ( 2\ell+1)} \sum_i \sum_m \left ( n_{\ell m}^{(i)} \right )^2
\end{equation}

These are respectively estimates of the noise pixel variance and of the noise power spectrum of our final map.

\subsection{Bias estimates}

The impact of the ILC bias is estimated by Monte-Carlo simulations on simulated data sets. The corresponding error is of order $2\%$ of the CMB.

\subsection{Practical implementation}

The practical implementation of this processing pipeline is made essentially using the {\tt octave} language (the free software version  of Matlab). The analysis is done in the framework of the pipeline tool developed in the ADAMIS team at the APC laboratory. This tool provides a flexible and convenient web interface for running our data analysis on simulations or real data with easy handling and tracing of the various pipeline options.\footnote{see http://www.apc.univ-paris7.fr/APC\_CS/Recherche/Adamis/ in the `outreach'  section} Single runs of the full pipeline require less than half an hour on a single processor of a standard desktop computer (dominated by harmonic transforms), whereas numerous pipelines on simulated data sets for Monte-Carlo are run on the ADAMIS 88-processor cluster, optimised for efficient I/O.

\section{Simulations}
\label{sec:simulation}

\subsection{The simulated data}

We start with a validation of our method on simulated data sets. For this experiment, synthetic observations of the sky emission are generated using the Planck Sky Model (PSM).
The PSM is a flexible software library, designed for simulating the total sky emission in the 10-1000 GHz frequency range, and developed as part of the foreground modelling activities of the Planck  working group on component separation (Planck WG2). Sky emission comprises galactic components of four origin (free-free, synchrotron, 
thermal dust, and spinning dust, with spectral emission laws for dust and synchrotron varying from pixel to pixel), CMB, kinetic and thermal SZ effect, and the emission from a population of galactic and extragalactic point sources which includes radio sources, infrared sources, and an infrared background. Although not perfect, the model sky is thought to be 
sufficiently representative of the complexity of the real sky emission for our simulations to be meaningful.

Sky maps are produced at WMAP central frequencies for the K, Ka, Q, V and W band, and convolved in harmonic space with approximate WMAP instrumental beams (Gaussian symmetric beams are used for these simulations). Uncorrelated, non-stationary Gaussian noise is added, with a pixel variance deduced from the WMAP sensitivity per channel and effective hit count. To mimic the subtraction of the brightest point sources detected by WMAP, we remove from the model sky, at each frequency, all sources with flux above 1 Jy (assuming they would have been detected, and can be subtracted from the data set). The 11 compact regions listed in table \ref{tab:sourcelist} however, being specific to the real sky, are not blanked for the simulations.

Although these simulations provide only an approximation of the real WMAP data sets, they are representative enough that the simulated data offer a component separation challenge close to that of the real data set. The IRIS map is used as part of the full set of data for the ILC implementation on simulations.

\begin{figure}[htb]
 \centering
 \resizebox{\hsize}{!}{\includegraphics[angle=90]{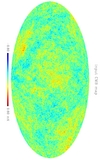}}
 \resizebox{\hsize}{!}{\includegraphics[angle=90]{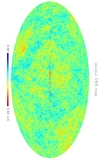}}
 \resizebox{\hsize}{!}{\includegraphics[angle=90]{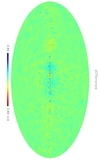}}
 \caption{Top: the simulated input CMB map. Middle: the reconstructed CMB. Bottom: the difference (output-input), displaying the residuals left by the method. All three maps share the same colour scale, and are at the resolution of the WMAP W channel.}
 \label{fig:cmb-simu-maps}
\end{figure}

\subsection{Results}

Figure~\ref{fig:cmb-simu-maps} shows the input simulated CMB, the output CMB, and the difference of the two for one particular simulation.  The reconstruction is visually good except in regions of local strong galactic emission (in the galactic ridge, for example). This is to be expected: not only the method can not remove foregrounds perfectly, but in addition the price to pay to remove foregrounds (even imperfectly) is more noise (because of sub-optimal weighting of the observations as far as noise contamination is concerned).

\begin{figure}[htb]
 \centering
 \resizebox{\hsize}{!}{\includegraphics{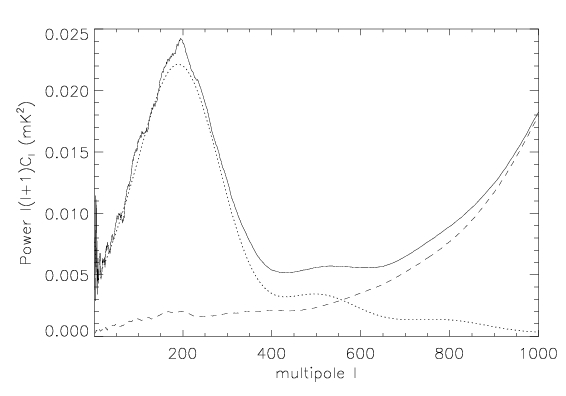}}
 \resizebox{\hsize}{!}{\includegraphics{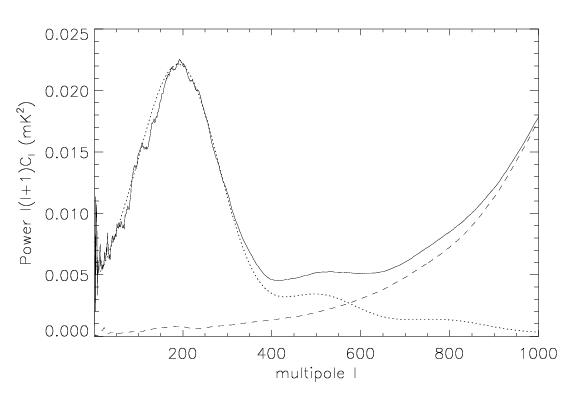}}
 \caption{Top: For simulated data sets, full sky power spectra of output CMB map (plain line), and of difference map (dashed line). The CMB model used in the simulation is over-plotted as a dotted line. Bottom: Same at high galactic latitude only (HGL region).}
 \label{fig:cmb-simu-spec}
\end{figure}

A more quantitative estimate of the level of contamination of the CMB map by foregrounds and noise is obtained by looking at power spectra. Figure~\ref{fig:cmb-simu-spec} shows the input simulated CMB power spectrum (dotted line), the spectrum of the output CMB (solid black line), and the spectrum of the map of residuals (difference between output and input, dashed line), both full sky (top panel) and in the HGL region (bottom panel). The angular power spectrum of the residual map is seen to be small as compared to the CMB power on large scales, the two being comparable at $\ell \simeq 500$. Noise dominates on smaller scales. The residuals due to the presence of galactic emission are seen to contribute power essentially below $\ell=400$, where the power of the difference map is seen to be slightly higher in the full sky power spectrum than in the HGL power spectrum (this is visible, in particular, at the top of the first acoustic peak).

\subsection{Bias}

As discussed in \ref{sec:ILC-bias}, we expect a (small) bias in the ILC map, due to empirical correlations between the CMB emission and contaminants (including noise and foregrounds). This is not particular to our approach, and is expected for any ILC method. For better characterisation of our output map, we evaluate the effect both theoretically (in appendix) and numerically.

Although a general analytic estimate of the bias is complicated, appendix \ref{app:bias} shows that (to first order at least and for reasonable assumptions about the CMB, foregrounds, and noise) the amplitude of the effect does not depend much on what the actual foregrounds are in detail, but is set essentially by the geometry of the domains considered (through a number of effective modes). It is then possible to estimate the amplitude of the effect by Monte-Carlo simulations on synthetic data sets resembling the actual WMAP observations.

\begin{figure}[htb]
 \centering
 \resizebox{\hsize}{!}{\includegraphics{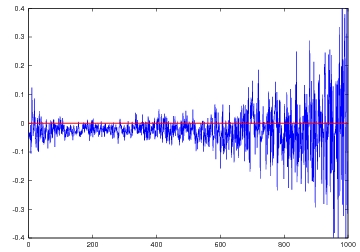}}
 \caption{The fractional bias in the ILC map as a function of $\ell$, for one single simulation. This figure is illustrative both of the amplitude of the effect, and on its variance for one single realisation. Bias, and standard deviation of the bias, are of the same order of magnitude for most of the $\ell$ range.}
 \label{fig:bias}
\end{figure}

Figure~\ref{fig:bias} illustrates an estimate of the bias $b(\ell)$ as a function of the harmonic mode, computed as a fractional error:
\begin{equation}
b(\ell) = \frac{\sum_m \left ( \widehat{a}_{\ell m} - {a}_{\ell m} \right ) {a}_{\ell m}^*}{\sum_m \left | {a}_{\ell m} \right |^2 }
\end{equation}
where ${a}_{\ell m}$ are the harmonic modes of the input CMB map, and $\widehat{a}_{\ell m}$ the harmonic modes of the output CMB map. The numerator in this equation computes the covariance of the residual map and the input map as a function of $\ell$, and the denominator is a normalisation factor. For an error uncorrelated with the input, $b(\ell)$ should be close to 0 on average. Analytical estimates of the effect (see appendix) suggest a bias of order 2\% for our implementation. The numerical estimate of Figure~\ref{fig:bias}, obtained as the average bias for 500 simulated data sets, is in good agreement with this prediction, with slight variations due to varying number of effective modes for different needlet scales. The average value of $b(\ell)$ between $\ell=2$ and 1000 on that simulation is about $-2.2$ \%.

\section{Application to WMAP data}
\label{sec:application}

\subsection{ILC Result}

We now turn to the description of the results obtained on the real
WMAP data sets.  In order to facilitate the comparison with existing
maps, we process independently one year, three year and five year
data, to obtain three CMB maps (hereafter NILC1, NILC3 and NILC5).
For each year, we use the beam estimates, noise level, and point
source catalogue provided with the corresponding release.

\begin{figure}[htb]
 \centering
 \resizebox{\hsize}{!}{\includegraphics[angle=0]{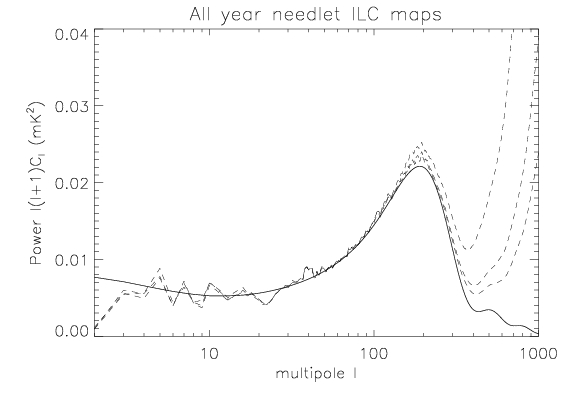}}
 \caption{Power spectra of the NILC map for one year, three year, and five year WMAP data}
 \label{fig:spec135-nomask}
\end{figure}

The improvement of CMB reconstruction with consecutive data releases is illustrated on Figure~\ref{fig:spec135-nomask}, which shows the full sky power spectra of the NILC CMB maps obtained with one year, three year, and five year WMAP data. The power spectra displayed are the raw power spectra of the output map, computed full sky, and smoothed with a variable window in $\ell$ of 10\% width. Whereas the lower part of the spectrum, cosmic variance limited and CMB dominated, does not change much, the high $\ell$ spectrum of the map, dominated by noise, decreases substantially with increasing observation time -- as expected. The excellent agreement at low $\ell$ (up to $\ell \simeq 300$) between the power spectra and the model is striking. The bumps in the spectrum due to the first and second acoustic peaks are cloearly visible on the five year map spectrum.

\begin{figure}[htb]
 \centering
 \resizebox{\hsize}{!}{\includegraphics[angle=90]{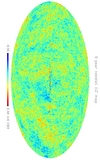}}
 \resizebox{\hsize}{!}{\includegraphics[angle=90]{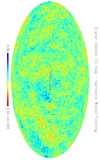}}
 \resizebox{\hsize}{!}{\includegraphics[angle=90]{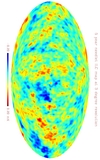}}
 \caption{Top: the NILC5 reconstructed WMAP CMB at the resolution of the W channel. Middle: the harmonic Wiener NILC5 CMB map. Bottom: The NILC5 CMB map at 3 degree resolution.}
 \label{fig:cmb-wmap-output}
\end{figure}

Our full-sky reconstructed CMB map for the five year observations, at the resolution of the W-channel, is displayed in the top panel of Figure~\ref{fig:cmb-wmap-output}.


\begin{figure}[htb]
 \centering
 \resizebox{\hsize}{!}{\includegraphics{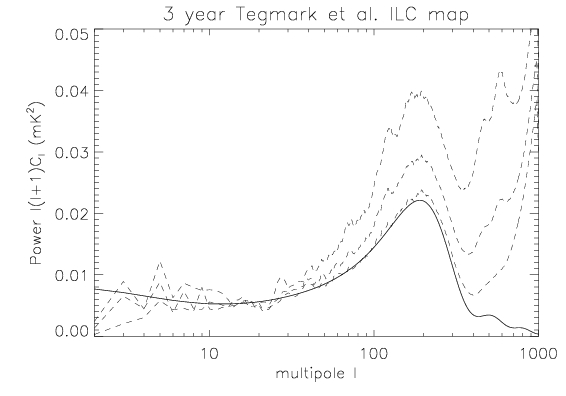}}
 \resizebox{\hsize}{!}{\includegraphics{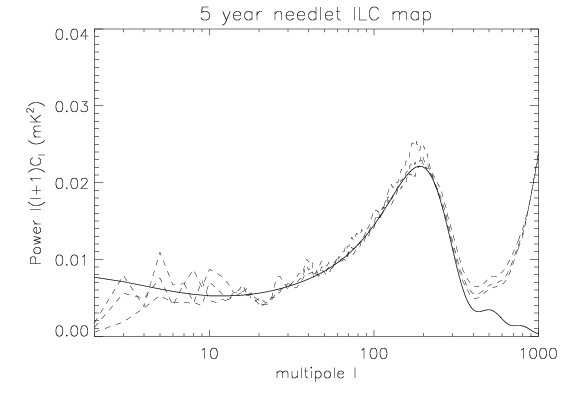}}
 \caption{Power spectrum of the reconstructed WMAP CMB map. For each of the two panels, the CMB best fit model is shown as a solid black line, and power spectra computed at  low galactic latitudes (using the LGL mask), on the full sky (no mask), and at high galactic latitudes only (HGL mask) are displayed as dashed lines. Note that the map spectra plotted here are directly those of the maps, without any correction for the beam. Top panel: ILC map of \citet{2003PhRvD..68l3523T} (TILC3). Bottom panel: this work, with five year WMAP data. The power spectrum of the needlet ILC CMB is significantly more homogeneous than the power spectrum of the TILC3 map. We interpret this difference as an indication that the TILC3 map is significantly more contaminated by residuals of galactic emission. Note the different scales of the $y$-axis, and the improvement on small scales, with a noise power of about 0.024~mK$^2$ at $\ell = 1000$ for the NILC5 map at high galactic latitude, instead of about 0.040~mK$^2$ for TILC3. As indicated by Figure~\ref{fig:spec135-nomask}, this is due essentially to the better quality of the five year release, since the NILC3 map also has a noise power spectrum of about 0.040~mK$^2$ at $\ell = 1000$.}
 \label{fig:cmb-spec-nilc-tilc}
\end{figure}

\section{Discussion}

\subsection{Comparison with other maps}
\label{sec:comparison}

A full comparison of our needlet ILC maps (for all data releases) with all the other available maps would be too long for the present discussion. Rather, we decide to compare our five year result only with the TILC3 map on small scales (choice is motivated by the fact that the TILC is the only other full sky high resolution map available), and with the EGS3 map on large scales. This is of particularly interest, as the EGS3 is the only map obtained with a method not based on the ILC, and also is a method specifically implemented  for the recovery of the largest scales.

\subsubsection{Comparison at the pixel level -- small scales}

\begin{figure*}[htb]
 \centering
 \resizebox{0.33\hsize}{!}{\includegraphics{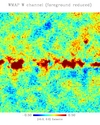}}
 \resizebox{0.33\hsize}{!}{\includegraphics{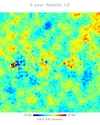}}
 \resizebox{0.33\hsize}{!}{\includegraphics{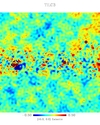}}\\
 \resizebox{0.33\hsize}{!}{\includegraphics{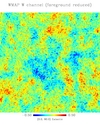}}
 \resizebox{0.33\hsize}{!}{\includegraphics{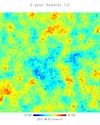}}
 \resizebox{0.33\hsize}{!}{\includegraphics{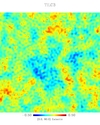}}
 \caption{CMB maps from WMAP three year data obtained with three techniques: left column: WMAP foreground reduced, W channel; middle column: our needlet ILC map; right column: the TILC map. The top line corresponds to a patch located in the galactic plane, centred around coordinates $(l,b)=(45^\circ,0^\circ)$. The bottom line shows the recovered CMB around the North Galactic Pole.}
 \label{fig:compare-map-small-scale}
\end{figure*}

On the smallest scales, we compare our needlet ILC map with the TILC and with the WMAP foreground-reduced W band.
Figure~\ref{fig:compare-map-small-scale} shows local regions of the foreground-reduced map, the NILC5 map, and the TILC3 map, in the galactic plane and at the galactic pole. Our needlet map is clearly less contaminated by galactic emission than the other two. At high galactic latitude, the NILC5 and TILC3 are visually comparable, while the foreground--reduced map appears to be more noisy, as expected.

The power spectrum of the output map for the five year data (NILC5 map), for three different sky coverages, is shown on the bottom panel of Figure~\ref{fig:cmb-spec-nilc-tilc}. On the same panel, we plot the angular power spectrum $C_\ell$ corresponding to the WMAP best fit model, corrected for the W-channel beam. On the top panel of the same figure, we show the same power spectrum estimates for the TILC3 map. This shows the improvement achieved by our method  close to the galactic plane. This improvement is due both to the needlet approach and to the use of the IRIS map to help with dust subtraction. As seen in Figure~\ref{fig:spec135-nomask}, the difference in quality between NILC5 and TILC3 cannot be explained solely by reduced noise (NILC5 and NILC3 being very close in quality for all scales except the smallest ones).

\subsubsection{Comparison at the pixel level -- large scales}

\begin{figure*}[htb]
 \centering
 \resizebox{0.49\hsize}{!}{\includegraphics[angle=90]{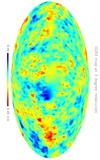}}
 \resizebox{0.49\hsize}{!}{\includegraphics[angle=90]{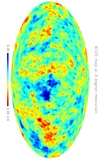}}
 \resizebox{0.49\hsize}{!}{\includegraphics[angle=90]{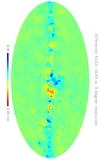}}
 \resizebox{0.49\hsize}{!}{\includegraphics[angle=90]{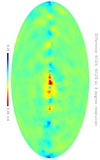}}
 \caption{Top left: The EGS3 map. Top right: The KILC5 map, smoothed at 3 degree resolution. Bottom left: The difference NILC5-EGS3; we see is a clear difference in the galactic plane with no particular structure, compatible with a smooth Gaussian field, where the EGS3 solution poorly estimates intermediate scales;  a patch in the difference map at the location of the Large Magellanic Cloud is clearly visible also. Bottom left: The difference NILC5-KILC5; a clear structure aligned with the galactic plane is clearly visible, with a big difference towards the galactic center.}
 \label{fig:compare-map-large-scales}
\end{figure*}

Figure~\ref{fig:compare-map-large-scales} gives a visual comparison of NILC5 (this work) and the EGS3 \citep{2007arXiv0709.1037E}, as well as of NILC5 and KILC5 \citep{2008arXiv0803.1394K}. On the top row, we display the EGS3 and KILC5 at a resolution of 3 degrees, and degraded to {\tt nside}$=64$. The bottom row shows the difference between our needlet ILC solution (displayed on the bottom panel of  Figure~\ref{fig:cmb-wmap-output}) and these two maps.

The most striking difference between the five year needlet ILC map and the EGS3 is in the galactic plane, where the EGS3 does not recover the intermediate angular scales. The difference, however, shows no clear particular structure, as expected if it is the random realisation of a Gaussian random field. It is thus probably essentially due to the difference between our CMB reconstruction on scales larger than 3 degrees, and the CMB on larger scales that can be inferred from the CMB reconstructed by Eriksen et al. outside of their galactic mask. At higher galactic latitude, the two maps are in good agreement, with no obvious feature which could be correlated to foregrounds or to the CMB itself, with the exception of a hot spot in the large Magellanic cloud (which might be residual emission of the LMC in our map, as \citet{2007arXiv0709.1037E} actually mask the centre of the LMC and obtain a solution in the direction of the LMC by extrapolation from nearby pixels).
Above 30 degree absolute galactic latitude, the RMS of the difference
between our 3 degree map and the EGS3 map is 5.7~$\mu$K. The two maps are in much better agreement than the EGS3 and the WMAP MEM model maps (see figure 3 in Eriksen et al.). Note however that theoretically, if there were no foregrounds in the WMAP data, the noise standard deviation $\sigma_n$ on a 3 degree map obtained by noise-weighted averaging using all WMAP channels would be about 3.2~$\mu$K for three year data. If instead we assumed that only the three highest frequency channels are free of foreground contamination, $\sigma_n$ would be 4.4~$\mu$K.

The difference between our map and the KILC5 map is more systematic with, in particular, stronger differences in the galactic plane, in spite of the fact that the two methods work on the exact same input data set. A careful visual inspection of the CMB maps themselves gives the impression that the KILC5 map is probably systematically negative towards the galactic central regions. There is, however, no secure way to be certain which map is best.

\subsection{Map characterisation}

\subsubsection{Beam}

\begin{figure}[htb]
 \centering
 \resizebox{\hsize}{!}{\includegraphics{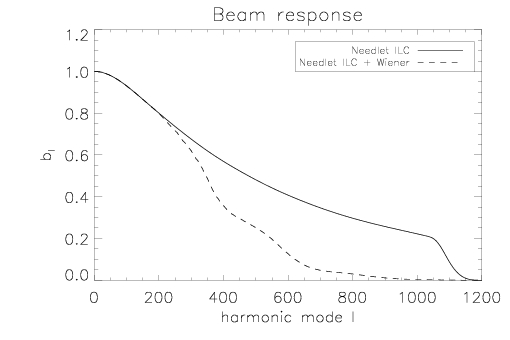}}
 \caption{Harmonic response of the beam of the NILC5 CMB map and of the Wiener-filtered version.}
 \label{fig:cmb-wmap-spec}
\end{figure}

The effective beam of the reconstructed maps are plotted in Figure~\ref{fig:cmb-wmap-spec}, for both the full resolution five year needlet ILC map, and for the Wiener-filtered version. The map has been reconstructed for the range $0 \leq \ell \leq 1200$, with a smooth transition of the response, between $\ell$ of 1024 and 1200, from the nominal W band beam value to 0. This smooth transition permits to avoid ringing effects which happen in case of a sharp cut--off in $\ell$. The ratio of the Wiener beam (dashed line) and ILC beam (solid line) gives a measure of the signal to noise ratio in each mode.

\subsubsection{Instrumental noise}

\begin{figure}[htb]
 \centering
 \resizebox{\hsize}{!}{\includegraphics[angle=0]{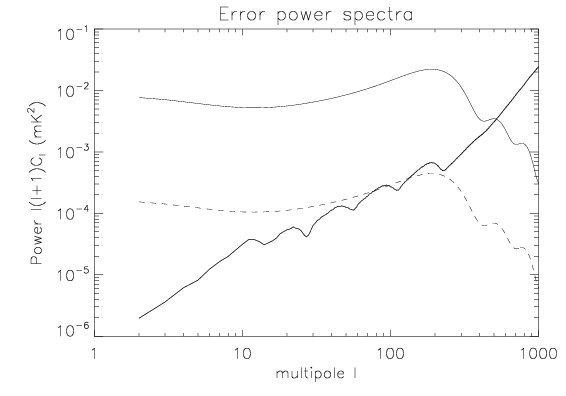}}
 \caption{Plot of the power spectrum of the noise (solid line). The spectrum for the WMAP best fit model is shown as a sold line also, for comparison. The dashed line is 2\% of the WMAP best fit $C_\ell$, indicative of the level of the expected ILC bias. The bias is seen to dominate on large scales. There is, however, little margin for improvement, as few independent modes (or needlet coefficients) are available on the largest scales.}
 \label{fig:plot_errors}
\end{figure}

\begin{figure}[htb]
 \centering
 \resizebox{\hsize}{!}{\includegraphics[angle=90]{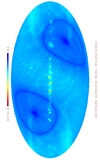}}
 \caption{Map of the standard deviation of the noise, per pixel at {\tt nside}=512, for the five year needlet ILC map at the resolution of the W channel. The adaptation of the filter to local contamination is obvious from the uniform noise level in large healpix pixels, in particular in the galactic plane. These large healpix pixels correspond to {\tt nside}=16, which is the size used for computing the average filter for the smallest scales. This illustrates the impact of the compromise between subtracting foregrounds and increasing the noise with the use of the lower frequency channels, which have poor sensitivity on small scales. }
 \label{fig:map_errors}
\end{figure}

Given the ILC filter computed on the real data set, the level and properties of the instrumental noise can be straightforwardly computed by applying the \emph{same} filter to simulated WMAP noise maps.
From 100 noise realisations, we compute the average full-sky noise power spectrum (Figure~\ref{fig:plot_errors}), as well as the noise standard deviation per pixel (figure \ref{fig:map_errors}).  Noise properties are not as simple as one may wish: the noise is non stationary, because of both the uneven sky coverage and of the localised processing. It is also somewhat correlated pixel-to-pixel, in particular close to the galactic plane. This is unavoidable, but fortunately our pipeline permits to make as many Monte-Carlo realisations of the instrumental noise as needed for any scientific study made using our needlet ILC map. Hundred such simulations are made available as part of our main data products.


\subsubsection{Foreground residuals}

More problematic is the evaluation of the contamination of the CMB map by foreground residuals. It requires prior information about the foregrounds, which the ILC method avoids using. An indication of the level of systematic uncertainty is obtained from the comparison of the various solutions (figure \ref{fig:stddev-map}). An other option consists in checking the contamination on simulations. Figures \ref{fig:cmb-simu-maps} and \ref{fig:cmb-simu-spec} give an idea of the expected contamination from such simulations. This, however, is good only as long as the simulations are representative enough. 

The comparison of the power spectrum of the output CMB map computed at high and low galactic latitude (figure \ref{fig:cmb-spec-nilc-tilc}), and a visualisation of the output CMP map at high and low galactic latitude (figure \ref{fig:compare-map-small-scale}) also give an idea of the amount of galactic residuals, but none of these estimates is fully satisfactory for careful CMB science.  This is, however, not particular to our map. No published CMB map is provided yet with a good estimate of foreground contamination. Although this, clearly, is not fully satisfactory, we leave further investigations on this question for future work.

\subsection{Final comments}

\subsubsection{Is our map optimal?}

In the present work, we have obtained a CMB map which has been shown to be significantly less contaminated by foregrounds and noise than the other existing maps obtained from WMAP data. A natural question is whether we can do even better. 

In the following, we outline where there is margin for improvement, and explain why we have stopped the analysis at its present state.

First of all, the present analysis uses only limited external information and data sets: WMAP point source detections, and the IRIS 100 micron map. It is likely that something could be gained by using additional observations to help constraining the galactic emission.

Second, there is a trade-off between localisation of the estimation of covariance matrices, and bias in the ILC.  The estimation of covariance matrices $\tR_x$ over sets of $N_p = 32 \times 32$ needlet coefficents results from a compromise which has been made based on some attempts with varying $N_p$ on simulations, but has not been optimised in any way. In addition, the optimal solution is probably different at high galactic latitudes, where weights given to different channels probably do not vary much and thus require less localisation, and at low galactic latitudes, where the complexity of galactic emission calls for more localisation. We have tried to use $N_p = 16 \times 16$ (more localised ILC filters, but more bias) and $N_p = 64 \times 64$ (less localised ILC filters, but less bias). Our choice of $N_p = 32 \times 32$ seems, on simulations, not worse than anything else (nor much better either). The bias error has been verified to remain below the reconstruction error due to the contribution of the noise for small scales, and below cosmic variance uncertainty for large scales (see figure \ref{fig:plot_errors}), and remains below an acceptable level of few percent.

Similarly, the choice of the spectral window functions used on this data set has not been the object of specific optimisation. At low $\ell$, we follow a ``dyadic'''scheme, where each window reaches an $\ell_{\rm max}$ about twice the previous one. Wide spectral window functions allow for more localisation in pixel space, but narrow window functions allow for more accuracy in the harmonic domain. At high $\ell$, because of the variation of the beams with $\ell$, the relative noise levels of the different channels change quite fast with $\ell$, which calls for more localisation in $\ell$ space. Here again, the optimal window functions are probably not the same at high and at low galactic latitudes. In practice, we chose a small number of bands to limit the number of harmonic transforms in the pipeline and allow reasonable localisation of the analysis.

The choice we have made results from the principle of simplicity. We have tried to devise a pipeline which depends as little as possible on external data, on priors, or on fine tuning. A very simple scheme has permitted to obtain a CMB map convincingly better than other maps available. This does not preclude any attempt at more optimisation for future work if needed.

\subsubsection{Why ILC and not ICA?}

It is certainly possible to tune our pipeline, changing some of its parameters. It would be possible also to use other methods than an ILC, for instance Independent Component Analysis (ICA) methods such as SMICA \citep{2003MNRAS.346.1089D,2008arXiv0803.1814C}, or more generally maximum likelihood methods fitting parametric models of the foregrounds on the data sets. Such methods extract information about the foregrounds from the data directly, possibly with the help of ancillary data sets, and use this information to clean the observations in some way.

In the present case however, it is not very likely that such attempts
would give much better results than what is obtained here, unless one
uses a very significant number of additional data sets and safe prior
information. Indeed, the WMAP data consists in five channels only,
from which a component separation method based on a meaningful model
of foreground emission needs to extract CMB, synchrotron, free-free,
spinning dust, thermal dust (i.e. five templates), and possibly also
point sources, and variations of the spectral indices of some of the
components -- not to mention special regions of galactic emission as
cold cores and H-II regions, nor particular objects as nearby resolved
galaxies or galaxy clusters. Any component separation method based on
the estimation of parameters for such a rich model would almost
certainly be confronted to indeterminacy issues. Methods based on a precise
model are expected to be effective when the data are very redundant as
compared with the number of ``parameters'' of the emission model, which
would not be the case here. Hence, the ILC is probably one of the best
approaches for doing component separation on WMAP data. It is not
surprising, then, that all methods producing full sky CMB maps from
WMAP, or nearly so, are implementing some variant of the ILC. Incidentally, ICA methods could benefit from the needlet framework.

\section{Conclusions}

In this paper, we have described a new approach for implementing CMB
extraction in WMAP data, using the ILC method on a needlet frame. Tests
on simulations show excellent performance of the method, thanks to
localisation both in pixel space and in harmonic space. Localisation
in pixel space allows the ILC weights to adapt themselves to local
conditions of foreground contamination and noise, whereas localisation
in harmonic space allows to favour foreground rejection on large scales
(where foregrounds dominate the total error) and noise rejection on
small scales (where foregrounds are negligible but where, after beam
deconvolution, the relative noise level between the various WMAP
channels varies a lot as a function of scale).  Needlets permit to
vary the weights smoothly on large scales, and rapidly on small
scales, which is not possible by cutting the sky in zones prior to any
processing.

As a further improvement to previous work on WMAP data, we include a dust template in the set of analysed observations. This is motivated by the fact that on the smallest scales, observed with reasonable signal to noise ratio by the W channel only, dust emission contributes a significant fraction of the total reconstruction of the map. Using the IRIS 100 micron map as an additional observation enables the ILC to reduce the final contamination by dust --thanks to correlations of dust emission between the W channel and the 100 micron map. Special care was also taken to subtract from the data, prior to making the ILC, a number of strong point sources. 

As discussed at length in the main text and in the appendix, the implementation of a filter (the ILC) based on empirical estimates of covariance matrices leads to a bias. This is not particular to our map, but is the case for any ILC map. We have estimated the level of this systematic effect, both analytically and numerically, to be at the level of a few per cent on all scales. Our simulation tool permits to make accurate estimates of the amplitude of the effect, if needed for any scientific exploitation of the NILC5 map.

The application of the method to WMAP one year and three year data (in
addition to five year data) permits to compare the needlet ILC solution
to previous work. Our map is seen to be at least as good as others on
large scales, while being significantly less contaminated by residual
foregrounds and noise than others on small scales, in particular in
the vicinity of the galactic plane.  The application of the method to
WMAP five year data yields a CMB map which we believe to be the cleanest
full sky map of the CMB to date. Contamination by noise, and the power
loss due to the use of the ILC method, are characterised by means of
Monte Carlo simulations. The map is available for download on the
ADAMIS web
site,\footnote{http://www.apc.univ-paris7.fr/APC\_CS/Recherche/Adamis/cmb\_wmap-en.php}
and can be used for a variety of science projects relying on accurate
maps of the CMB.

\begin{acknowledgements}
  The ADAMIS team at APC has been partly supported by the Astro-Map and Cosmostat ACI grants
  of the French ministry of research, for the development of innovative CMB data analysis methods.
  We acknowledge the use of the Legacy Archive for Microwave Background Data
  Analysis (LAMBDA). Support for LAMBDA is provided by the NASA Office of Space Science.
  The results in this paper have been derived using the
  HEALPix package \citep{2005ApJ...622..759G}.
  The authors acknowledge the use of the Planck Sky Model,\footnote{http://www.apc.univ-paris7.fr/APC\_CS/Recherche/Adamis/PSM/psky-en.php} developed by the Planck working group on component separation, for making the simulations used in this work.
  Our pipeline is mostly implemented in octave (www.octave.org).
\end{acknowledgements}

\bibliographystyle{aa} 
\bibliography{biblio} 

\clearpage
\appendix

\section{Derivation of the ILC bias}
\label{app:bias}

In this appendix, we compute the error made after CMB reconstruction
with the ILC and some of its statistical properties. In particular, we
derive the correlation of the error with the true CMB signal, which
yields a non unit effective ``response'' of the ILC filter -- and hence
a {\emph{bias}} in the reconstructed map and in the CMB power spectrum computed
from it. This bias has to be accounted for properly for further uses
of the reconstructed CMB map.

We model the data as:
\begin{equation}
\vx_p = \va s_p + \vn_p
\end{equation}
where $s_p$ are the coefficients of the map of interest over some domain $\mathcal{D}$ (e.g. needlet coefficients of the CMB map for a given scale and a given patch of the sky, or pixel values in a certain region of the sky, or values of the harmonic coefficients in some band of $\ell$). $p$ indexes the coefficient (i.e. pixel, or harmonic mode, or needlet coefficient). $\vx_p$ are the observations of coefficient $p$ for the set of available observed maps, and $\vn_p$ the corresponding ``noise'' terms (including foreground contaminants).

The ILC is best applied over domains of $p$ where all coefficient have
(near) uniform expected signal and noise properties, so that the ILC
weights are optimal simultaneously for all $p$.  In particular, the
RMS values of all maps do not depend (much) on $p$ in a given domain.
Hence, we concentrate on one given domain of $p$ for which we assume
that the sequences $s_p$ and $\vn_p$ are independent, Gaussian random
variables with distribution $\mathcal{N}(0,\sigma_s^2)$ and
$\mathcal{N}(0,\tR_n)$, with $\sigma_s^2$ the variance of $s_p$ (the
CMB) and $\tR_n$ the covariance matrix of the noise (including
foregrounds).

The ILC estimate of $s_p$ in domain $\mathcal{D}$ is given, for all
$p$, by
\begin{equation}
  \hat{s}_p = \frac{\va^t \, {\widehat{\tens{R}}_x}^{-1}}{\va^t \, {\widehat{\tens{R}}_x}^{-1} \, \vec{a}} \, \vx_p
\end{equation}
where $\widehat{\tens{R}}_x$ is an estimate of the covariance matrix
of the observations, obtained as:
\begin{eqnarray}
  \widehat{\tens{R}}_x 
  & = & \frac{1}{N_p} \sum_p \vx_p \vx_p^t \nonumber \\
  & = & \frac{1}{N_p} \sum_p (\va s_p + \vn_p )(\va s_p + \vn_p )^t
  \label{eq:empirical-covariance-matrix}
\end{eqnarray}
with $N_p$ the number of coefficients in domain $\mathcal{D}$.  In the
limit of large $N_p$, $\widehat{\tens{R}}_x $ approaches its
expectation (ensemble average) value $\mathrm{E}(\widehat{\tens{R}}_x)
= {\tens{R}}_x$. For finite $N_p$, we have instead
\begin{equation}
\widehat{\tens{R}}_x = \tens{R}_x + \tens{\Delta}_x
\label{eq:correction-term}
\end{equation}
where $\tens{\Delta}_x$ is a correction term corresponding to the
departure of the empirical correlation from its ensemble average due
to the finite sample size $N_p$.  From now on, we assume that $N_p$ is
large enough that this correction is small: we investigate effects at
first order in $1/N_p$.

\subsection{First order expansion of the reconstruction error}
\label{sec:firstorderbias}

We are interested in the reconstruction error:
\begin{equation}
  d_p = \hat{s}_p - s_p
\end{equation}
The first and second moments of $d_p$, i.e. the mean value
$\mathrm{E}(d_p)$ of the reconstruction error, as well as its variance
$\mathrm{E}({d_p}^2)$, are of particular interest for the
interpretation of the reconstructed map.  In particular, we have
\begin{equation}
  \mathrm{E}({\hat{s}_p}^2) = \mathrm{E}({s_p}^2) + \mathrm{E}({d_p}^2) + 2\mathrm{E}(s_p d_p)
\label{eq:powerspec-with-bias}
\end{equation}
In our case, $\mathrm{E}({{\hat{s}}_p}^2)$ can be used to estimate
$\mathrm{E}({s_p}^2)$.  In the case where domain $\mathcal{D}$ is an
harmonic domain, $p$ indexes harmonic modes $(\ell,m)$, and
$\mathrm{E}({s_p}^2)$ is a term of the CMB power spectrum.  In our
needlet approach, $\mathrm{E}({s_p}^2)$ is also directly connected to
the CMB power spectrum. For this reason, it is important to
characterise in the best way we can the ``noise bias''
$\mathrm{E}({d_p}^2)$ and the covariance of the error with the CMB,
$\mathrm{E}(s_p d_p)$.

The ILC being constructed so that the response to the signal of interest is unity, only the filtered noise term contributes to the error $d_p$, which can then be written as:
\begin{eqnarray}
d_p & = & \frac{\va^t \, {\widehat{\tens{R}}_x}^{-1}}{\va^t \, {\widehat{\tens{R}}_x}^{-1} \, \vec{a}} \, \vn_p \\
	& = & \frac{\va^t \, \left [ \tens{R}_x + \tens{\Delta}_x \right ] ^{-1}}{\va^t \, \left [ \tens{R}_x + \tens{\Delta}_x \right ] ^{-1} \, \vec{a}} \, \vn_p
\end{eqnarray}
where 
$\tens{\Delta}_x$ is a small perturbation to $\tens{R}_x$. 
We use the first order expansion:
\begin{equation}
\left [ \tens{R} + \tens{\Delta}_x \right ] ^{-1} \simeq  \tens{R} ^{-1} - \tens{R} ^{-1} \, \tens{\Delta}_x \, \tens{R} ^{-1}
\end{equation}
which yields
\begin{equation}
d_p = \frac{\va^t \, \left [ \tens{R}_x ^{-1} - \tens{R}_x ^{-1}  \tens{\Delta}_x  {\tens{R}}_x ^{-1} \right ] \, \vn_p}
	{\va^t \, \left [ {\tens{R}}_x ^{-1} - {\tens{R}}_x ^{-1}  \tens{\Delta}_x  {\tens{R}}_x ^{-1}  \right ] \, \vec{a}} 
\end{equation}
Writing:
\begin{eqnarray}
\frac{1}{\va^t \, \left [ {\tens{R}}_x ^{-1} - {\tens{R}}_x ^{-1}  \tens{\Delta}_x {\tens{R}}_x ^{-1}  \right ] \, \vec{a}} 
	& = & \frac{1}{\va^t {\tens{R}}_x ^{-1} \va} \, \frac{1}{(1-\epsilon)} \nonumber\\
	& \simeq & \frac{1}{\va^t {\tens{R}}_x ^{-1} \va} \, (1+\epsilon) \nonumber
\end{eqnarray}
where $\epsilon$ is 
\begin{equation}
\epsilon = \frac{\va^t {\tens{R}}_x ^{-1}  \tens{\Delta}_x  {\tens{R}}_x ^{-1} \va}   {\va^t {\tens{R}}_x ^{-1} \va} \nonumber
\end{equation}
we get
\begin{equation}
d_p \simeq \frac{\va^t \, \left [ {\tens{R}}_x ^{-1} - {\tens{R}}_x ^{-1} \tens{\Delta}_x {\tens{R}}_x ^{-1} \right ] \, \vn_p}
	{\va^t {\tens{R}}_x ^{-1} \va} \, \left ( {1+\epsilon} \right ) 
\end{equation}
Keeping only first order terms in $\left ( \tens{\Delta}_x \right )$ yields:
\begin{eqnarray}
d_p  & =  & \frac{\va^t {\tens{R}}_x ^{-1} \vn_p}{\va^t {\tens{R}}_x ^{-1} \va}  -  \,
	\frac{\va^t {\tens{R}}_x ^{-1} \tens{\Delta}_x {\tens{R}}_x ^{-1} \vn_p}{\va^t {\tens{R}}_x ^{-1} \va} \nonumber \\
	& & +   \, \frac{\left [ \va^t {\tens{R}}_x ^{-1} \vn_p \right ] \left [ \va^t {\tens{R}}_x ^{-1} \tens{\Delta}_x {\tens{R}}_x ^{-1} \va \right ]}{ \left [ \va^t {\tens{R}}_x ^{-1} \va 
	\right ]^2}
	\label{eq:diff-dp}
\end{eqnarray}
The first term on the right hand side, proportional to $\vn$, is the ``ideal'' ILC error, i.e. the error we would get if we knew perfectly the ``true'' covariance matrix $\tens{R}_x$ of the observations. The second and third terms, proportional to 
$\tens{\Delta}_x$, are corrections due to the fact that this covariance matrix is actually obtained empirically from the observation themselves.

From equations \ref{eq:empirical-covariance-matrix} and \ref{eq:correction-term}, we can write $\tens{\Delta}_x$ in the form:
\begin{equation}
\tens{\Delta}_x = \delta_s \va\va^t + \tens{\Delta}_n + \widehat{\tens{C}}
\end{equation}
where 
\begin{equation}
\delta_s  =  \widehat{\sigma}_s^2 - \sigma_s^2 \nonumber
\end{equation}
\begin{equation}
\tens{\Delta}_n  =  \widehat{\tR}_n - {\tR}_n \nonumber
\end{equation}
\begin{equation}
\widehat{\tens{C}}  =  \frac{1}{N_p} \sum_q (\vn_q s_q \va^t + \va s_q \vn_q^t ) \nonumber
\end{equation}
These three quantities correspond respectively, in pixel (or mode, or
needlet coefficient) $p$, to the uncertainty in CMB variance estimates
due to ``cosmic'' (or sample) variance, to the error in the estimation
of the ``noise'' covariance matrix alone (if maps of noise+foregrounds
alone were available), and to a cross term, originating from the
empirical covariance between CMB and noise due the finite sample size
$N_p$.

The two last terms (small correction terms) in equation \ref{eq:diff-dp}, being proportional to $\tens{\Delta}_x$, can be decomposed each into the sum of three terms, proportional to $(\delta_s \va\va^t)$, $\tens{\Delta}_n$, and $\widehat{\tens{C}}$ respectively. The signal and noise realisations enter the $(\delta_s \va\va^t)$ term as products of terms of the form $(\va \va^t s_q s_q \vn_p)$ only, the $\tens{\Delta}_n$ term as products of terms of the form $(\vn_q \vn_q^t \vn_p)$. On the contrary, signal and noise realisations enter the $\widehat{\tens{C}}$ term as the product of terms in the form $(\va s_q \vn_q^t \vn_p)$, i.e. products of $s$ and second power of $\vn$.
Index $q$ runs over domain $\mathcal{D}$.

Assuming that $s$ and $\vn$ are centred variables, the mean value of the error is immediately seen to vanish:
\begin{equation}
\mathrm{E}(d_p) = 0
\end{equation}
The main contribution to the variance comes from the first term on the right hand side of eq.~\ref{eq:diff-dp}. The second and third terms are small corrections to this variance estimate, so that to first order, we get:
\begin{equation}
\mathrm{E}(d_p^2) \simeq \frac{\va^t {\tens{R}}_x ^{-1} {\tens{R}}_n {\tens{R}}_x ^{-1} \va}{ \left [ \va^t {\tens{R}}_x ^{-1} \va \right ]^2}
\end{equation}
Recalling that $\tR_x = \tR_n + \sigma^2 \va \va^t$, where $\sigma^2$ is the variance of the CMB, and making use of the inversion formula:
\begin{equation}
  \left [ \tR_n + \sigma^2 \va \va^t \right ] ^{-1}
  = \tR_n^{-1} - \sigma^2 \frac{\tR_n^{-1} \va \va^t \tR_n^{-1}}{1+ \sigma^2 \va^t \tR_n^{-1} \va}
\end{equation}
we finally obtain:
\begin{equation}
  \mathrm{E}(d_p^2) \simeq \frac{1}{ \left [ \va^t {\tens{R}}_n ^{-1} \va \right ]} .
\end{equation}

The most interesting terms are those connecting the error to the signal of interest, $\mathrm{E}(s_p d_p)$, which is necessary to compute the power spectrum of the output map according to \ref{eq:powerspec-with-bias}.\footnote{We warn the reader that some authors fail to make a clear distinction between the statistical (ensemble average) correlation, which is a deterministic quantity, and the ``empirical correlations'' computed, assuming some kind of ergodicity, as averages over finite sets of samples as in equation \ref{eq:empirical-covariance-matrix}.}

As mentioned previously, under the assumption that the signal of interest is not correlated to the noise and the foregrounds, the first term (main term) of the r.h.s. of equation~\ref{eq:diff-dp} does not give rise to multiplicative errors (or correlation of $d_p$ with $s_p$). Similarly, the corrective term proportional to $\delta_s \va\va^t + \tens{\Delta}_n$, multiplied by $s$, gives terms which contain an odd power of $s$ and an odd power of $\vn$, and does not give rise to correlations. This assumption is correct, to excellent accuracy, when the signal of interest is CMB anisotropies.\footnote{Certainly the CMB is not correlated to galactic components. Small correlations with large scale structure, and hence with SZ effect and emission from outer galaxies, may exist because of the integrated Sachs-Wolfe effect. We neglect this effect in the present discussion.} We are left with:

\begin{eqnarray}
\mathrm{E}(s_p d_p)  =  & & \mathrm{E} \left (  \sum_q \frac{s_p s_q}{N_p} \frac{\left [ \va^t {\tens{R}}_x ^{-1} \vn_p \right ] \left [ \va^t {\tens{R}}_x ^{-1} (\vn_q \va^t + \va \vn_q^t ) {\tens{R}}_x ^{-1} \va \right ]}{ \left [ \va^t {\tens{R}}_x ^{-1} \va \right ]^2} \right ) \nonumber \\
& - & \mathrm{E} \left ( \sum_q \frac{s_p s_q}{N_p} \frac{\va^t {\tens{R}}_x ^{-1}  (\vn_q \va^t + \va \vn_q^t )  {\tens{R}}_x ^{-1} \vn_p}{\va^t {\tens{R}}_x ^{-1} \va}  \right )
\end{eqnarray}
Multiplying the numerator and denominator of the second term by $\va^t {\tens{R}}_x ^{-1} \va$ and expanding  numerators, two terms cancel and two remain.
f we assume in addition that signal and/or noise coefficients are independent, i.e. $\mathrm{E}(s_p s_q) = \sigma_s^2 \delta_{qp}$ and/or $\mathrm{E}(\vn_p \vn_q^t) = \tR_n \delta_{qp}$, only the ${pp}$ term is non vanishing, and we get
\begin{equation}
\mathrm{E}(s_p d_p) =  \frac{\sigma_s^2}{N_p} \, \left ( \frac{  \mathrm{E} \left ( (\va^t {\tens{R}}_x ^{-1} \vn_p)^2 \right ) 
	- (\va^t {\tens{R}}_x ^{-1} \va ) \,\mathrm{E} \left ( \vn_p^t {\tens{R}}_x ^{-1} \vn_p \right )}{ \left [ \va^t {\tens{R}}_x ^{-1} \va \right ]} \right )
	\label{eq:almost-there}
\end{equation}
We compute
\begin{eqnarray}
\mathrm{E} \left ( (\va^t {\tens{R}}_x ^{-1} \vn_p)^2 \right ) & = & \va^t {\tens{R}}_x ^{-1} {\tens{R}}_n {\tens{R}}_x ^{-1} \va \nonumber \\
	& = & \va^t {\tens{R}}_x ^{-1} \left [ \tens{R}_x - \sigma_s^2 \va \va^t \right ] {\tens{R}}_x ^{-1} \va \nonumber \\
	& = & \left [ \va^t {\tens{R}}_x ^{-1} \va \right ] \left [ 1 - \sigma_s^2 \, \va^t {\tens{R}}_x ^{-1} \va \right ]
	\label{eq:E1}
\end{eqnarray}
and
\begin{eqnarray}
\mathrm{E} \left ( \vn_p^t {\tens{R}}_x ^{-1} \vn_p \right ) & = & \mathrm{Tr} \left ( {\tens{R}}_x ^{-1} {\tens{R}}_n \right ) \nonumber \\
	& = &  \mathrm{Tr} \left ( {\tens{R}}_x ^{-1} \left [ \tens{R}_x - \sigma_s^2 \va \va^t \right ]  \right ) \nonumber \\
	& = &  \mathrm{Tr} \left ( \tens{Id} \right ) - \sigma_s^2 \, \mathrm{Tr} \left ( {\tens{R}}_x ^{-1}  \va \va^t \right ) \nonumber \\
	& = &  m - \sigma_s^2  \left ( \va^t {\tens{R}}_x ^{-1}  \va \right )
	\label{eq:E2}
\end{eqnarray}
where $m$ is the number of channels used for the ILC (here, 5 WMAP channels + 1 IRIS map, for a total of 6). Substituting the results of equations \ref{eq:E1} and \ref{eq:E2} into equation \ref{eq:almost-there}, we get the simple final result:
\begin{equation}
\mathrm{E}(s_p d_p)  = \frac{\sigma_{s}^2 (1-m)}{N_p}
\label{eq:result}
\end{equation}
The error in the reconstructed CMB map comprises a term proportional (on average) to the CMB. In our application, $m=6$ and $N_p = 1024$, so that if indeed all needlet coefficients were independent, the amplitude of the effect should be $\mathrm{E}(s_p d_p) \simeq 5 \times 10^{-3} \sigma_s^2$, i.e. a bias of about half a percent in the CMB reconstruction.

\subsection{A geometric interpretation}

Although allowing the statistical derivation of the (anti-) correlation of the reconstruction error with the original CMB, the above derivation is not very illuminating about the mechanism giving rise to this CMB power loss. A geometrical reasoning gives better insight on what is actually going on.

For a given data set, the ILC works on one single realisation of all random fields. For an independent implementation of the ILC on $N_p$ pixels (or modes, or needlet coefficients) of the observations, each data set is represented by a vector in a $N_p$-dimensional vector space $W$. The CMB $\vs$, the observation $\vx_i$ for each channel $i$, and each of the noise realisations $\vn_i$ (including foregrounds) are elements of $W$.

The collection of vectors $\vn_i$ defines an $m$-dimensional subspace $V$ of $W$. This is true irrespective of the nature of the foregrounds: indeed, although in principle vectors $\vn_i$ could be linearly dependent, this happens in practice with vanishing probability (in particular if the observations are noisy).

Vector space $W$ can thus be decomposed in two orthogonal subspaces, $U$ and $V$, where $V$ is the $m$-dimensional sub-space spanned by all vectors $\vn_i$, and $U=V^\perp$ is a $(N_p-m)$ dimensional vector space. The CMB itself can be decomposed into two components, one in $U$, and one in $V$:
\begin{equation}
\vs = \vs_U + \vs_V
\end{equation}
where $\vs_U$ is the orthogonal projection of $\vs$ onto $U$, and $\vs_V$ its orthogonal projection onto $V$.

\begin{figure}[htb]
 \centering
 \resizebox{\hsize}{!}{\includegraphics[angle=0]{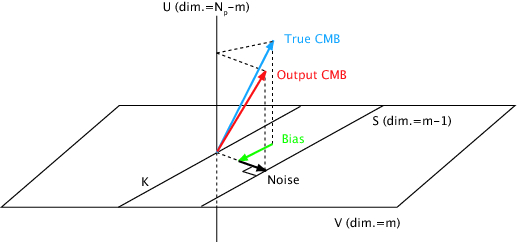}}
 \caption{Geometric illustration of the ILC bias. The true CMB (blue) and the output CMB (red) are elements of $W=U \oplus V$. The difference between the two can be decomposed into the sum of two elements of $V$: a bias (green), which is an element of $K$, and a noise contribution (black), which is an element of $H$, the orthogonal of $K$ in $V$.}
 \label{fig:biais-geom}
\end{figure}

What happens when the ILC is made is the following: we look for weights $w_i$ for all channels, that minimise the variance of the reconstructed map, i.e. minimise the norm of vector $\widehat \vs = \sum_i w_i \vx_i$, under the constraint that $\sum_i w_i = 1$. We have:
\begin{eqnarray}
\widehat \vs & = & \vs_U + \left ( \vs_V + \sum_i w_i \vn_i \right ) \nonumber \\
		    & = & \widehat \vs_U + \widehat \vs_V
\end{eqnarray}
where the first term is a vector of $U$ and the second term a vector of $V$, and the second line of the equation defines $\widehat \vs_U$ and $\widehat \vs_V$. Since these two subspaces of $W$ are orthogonal, the norm of $\widehat \vs$ is the sum of the norms of the two vectors $\widehat \vs_U$ and $\widehat \vs_V$. The norm of $\widehat \vs$ thus depends on $w_i$ only through the norm of the projection of $\widehat \vs$ on subspace $V$.

The noise contribution to $\widehat \vs$ appears as a linear combination of vectors $\vn_i$. For varying values of $w_i$ such that $\sum_i w_i = 1$, this linear combination spans an \emph{affine} subspace $S$ of $V$. $S$ is of dimension $m-1$ (an hyperplane). Defining $K$ as the vector subspace of $V$ spanned by linear combinations $\sum_i w_i \vx_i$ such that $\sum_i w_i = 0$, we obtain $S$ as:
\begin{equation}
S = \vp + K
\end{equation}
where $\vp$ is any element of $S$.

We note that the vector subspace $K$ \emph{depends only on noise realisations, and not on} $\vs$ \emph{nor on the final ILC weights} (the latter only defining a single point on $S$ -- and on $K$ by orthogonal projection). Hence, the direction of the one-dimensional vector subspace $H$ of $V$ orthogonal to $K$ is also independent of $\vs$  and of the final ILC weights.

For any element $\widehat \vs_V = \vs_V + \sum_i w_i \vn_i$ of affine space $S$ the norm of $\widehat \vs_V$ is the sum of the norms of its projections onto $K$ and $H$. 
Allowing weights $w_i$ to vary, vector $\widehat \vs_V$ spans $S$, and hence only the norm of the projection of $\widehat \vs_V$ onto $K$ varies (and not its projection on $H$). The minimum is reached when the projection of of $\widehat \vs_V$ onto $K$ vanishes. When this happens, the ILC has cancelled completely the linear combination of projections of $\vs$ and $\vn_i$ onto the $m-1$-dimensional space $K$, and left untouched the projections of $\vs$ and $\vn_i$ onto the $N_p-m+1$ dimensional vector space $U \oplus H$. 

Assuming the CMB to be Gaussian and statistically isotropic, its
coefficients in any orthogonal basis are Gaussian distributed random
variables with variance $\sigma^2/N_p$ (since the sum must have total
variance $\sigma^2$).  It follows straightforwardly that the
correlation of the recovered CMB map with the input CMB map is
$(N_p-m+1)/N_p$, and that the ``bias'' is due to the loss of $m-1$ modes
of the original CMB, which have been unlucky enough to ``live'' in the
$(m-1)$ dimensional space $K$.

\subsection{Comment on coefficient independence}

The above derivation in Section~\ref{sec:firstorderbias} assumes the independence of coefficients $\vn_p$ and/or of coefficients $s_p$, i.e. $\mathrm{E}(\vn_p \vn_q^t) = \tR_n \delta_{qp}$ and/or $\mathrm{E}(s_p s_q) = \sigma_s^2 \delta_{qp}$. When this assumption does not hold, we have:
\begin{eqnarray}
\mathrm{E}(s_p d_p) & = & \frac{1}{N_p} \, \sum_q \, \mathrm{E} \left ( s_p s_q \, \frac{  (\va^t {\tens{R}}_x ^{-1} \vn_p) (\va^t {\tens{R}}_x ^{-1} \vn_q) }
					{\left [ \va^t {\tens{R}}_x ^{-1} \va \right ]} \right ) \nonumber \\
	& & - \frac{1}{N_p} \, \sum_q \, \mathrm{E} \left ( s_p s_q \, \left ( \vn_p^t {\tens{R}}_x ^{-1} \vn_q \right ) \right )
\end{eqnarray}
Assuming that the noise and the CMB are independent, we have:
\begin{eqnarray}
\mathrm{E}(s_p d_p) & = & \frac{1}{N_p} \, \sum_q \, \mathrm{E} \left ( s_p s_q \right ) \, \frac{ \mathrm{E} \left (   (\va^t {\tens{R}}_x ^{-1} \vn_p) (\va^t {\tens{R}}_x ^{-1} \vn_q) 
	\right ) } 
					{\left [ \va^t {\tens{R}}_x ^{-1} \va \right ]}  
\nonumber \\
	& & - \frac{1}{N_p} \, \sum_q \, \mathrm{E} \left ( s_p s_q \right ) \, \mathrm{E} \left ( \vn_p^t {\tens{R}}_x ^{-1} \vn_q \right )
	\label{eq:Espdp-general}
\end{eqnarray}
where
\begin{equation}
\mathrm{E} \left ( (\va^t {\tens{R}}_x ^{-1} \vn_p)(\va^t {\tens{R}}_x ^{-1} \vn_q) \right ) =  \va^t {\tens{R}}_x ^{-1} \mathrm{E} \left ( \vn_p \vn_q^t  \right ) {\tens{R}}_x ^{-1} \va 	\label{eq:E1pq}
\end{equation}
and
\begin{equation}
\mathrm{E} \left ( \vn_p^t {\tens{R}}_x ^{-1} \vn_q \right ) = \mathrm{Tr} \left ( {\tens{R}}_x ^{-1} \mathrm{E} \left ( \vn_p \vn_q^t   \right ) \right ) 
	\label{eq:E2pq}
\end{equation}

When $p$ and $q$ index needlet coefficients as in the present work, we have:
\begin{equation}
\mathrm{E} \left ( s_p s_q  \right ) = \sum_\ell \frac{2\ell + 1}{N_{\rm tot}}  h_\ell^2 C_\ell P_\ell(\cos \theta_{qp})
\end{equation}
where $s_p$ and  $s_q$ are needlet coefficients of the CMB map, evaluated at two different points $p$ and $q$, $C_\ell$ is the angular power spectrum of the CMB, $\theta_{qp}$ is the angle between $q$ and $p$, and $N_{\rm tot}$ is the total number of pixels of the needlet coefficient map. For noise maps (including foregrounds), which are not stationary Gaussian random fields on the sphere, the analogous formula is just an approximation, which can be written as:

\begin{equation}
\mathrm{E} \left ( \vn_p \vn_q^t  \right ) \simeq \sum_\ell \frac{2\ell + 1}{N_{\rm tot}}  h_\ell^2 \, \tR_n(\ell) \, P_\ell(\cos \theta_{qp})
\end{equation}
\noindent where $\vn_p$ and  $\vn_q$ are needlet coefficients of all noise maps, evaluated at two different points $p$ and $q$, $\tR_n(\ell)$ is the covariance of the noise needlet coefficients (an $m \times m$ matrix for each $\ell$), and $\theta_{qp}$ and $N_{\rm tot}$ are defined as above.

Assuming that neither the noise level, nor the CMB power, do vary much over the spectral window $h_\ell$, $\tR_n(\ell)$ and $C_\ell$ are approximately independent of $\ell$, and can be taken out of the integral. We get:
\begin{equation}
\mathrm{E} \left ( s_p s_q \right ) = C_\ell \, k(\theta_{qp})
\end{equation}
and 
\begin{equation}
\mathrm{E} \left ( \vn_p \vn_q^t  \right ) = \tR_n \, k(\theta_{qp})
\end{equation}
with 
\begin{equation}
k(\theta_{qp}) =  \sum_\ell \frac{2\ell + 1}{N_{\rm tot}}  h_\ell^2 P_\ell(\cos \theta_{qp})
\end{equation}
Hence, plugging this result, together with equations \ref{eq:E1pq} and \ref{eq:E2pq}, into equation \ref{eq:Espdp-general} we get:

\begin{equation}
\mathrm{E}(s_p d_p) = \frac{(1-m)}{N_p} \, C_\ell \, \sum_q k^2(\theta_{qp})
\end{equation}
Finally, noting that $\sigma_s^2 = C_\ell k(0)$, we get 
\begin{equation}
\mathrm{E}(s_p d_p) = \frac{(1-m)}{N_p} \, \frac{\sigma_s^2}{k(0)} \, \sum_q k^2(\theta_{qp})
\label{eq:resultpq}
\end{equation}
Equation \ref{eq:resultpq} is the exact same as equation \ref{eq:result}, except for a coefficient, which measures the correlation between signal and noise coefficients $p$ and coefficients $q$ in domain $\mathcal{D}$. In particular, the result is, again, independent of $\tR_n$.

Hence, we define an {\emph{effective number of modes}}, $N_p^{\rm eff} = {N_p}/{\alpha}$, where
\begin{equation}
\alpha = \frac{\sum_q k^2(\theta_{qp})}{k(0)}
\end{equation}
and we get
\begin{equation}
\mathrm{E}(d_p s_p)  = \frac{\sigma_{p}^2 (1-m)}{N_p^{\rm eff}}
\end{equation}
An approximation (and upper bound) for $\alpha$ is easily obtained in the special case where $h_\ell$ is a square spectral window, and when domains $\mathcal{D}$ over which the ILC is implemented are small regions of the sky, so that $h_\ell^2 = 1$, and $P_\ell(\cos \theta_{qp}) \simeq 1$. In this case, we have:
\begin{equation}
k(\theta_{qp}) \simeq \frac{1}{N_{\rm tot}} \left ( (\ell_{\rm max}+1)^2 - \ell_{\rm min}^2 \right )
\end{equation}
and we have
\begin{equation}
N_p^{\rm eff} \simeq \frac{N_p}{N_{\rm tot}} \left ( (\ell_{\rm max}+1)^2 - \ell_{\rm min}^2 \right )
\end{equation}
We note that $\left ( (\ell_{\rm max}+1)^2 - \ell_{\rm min}^2 \right)$
simply is the number of harmonic coefficients selected by the spectral
window $h_\ell$, and that ${N_p}/{N_{\rm tot}} = f_{\rm sky}$ is a
coefficient which takes into account the effect of partial sky
coverage for the (local) calculation of the statistics of the data
set, well in line with $N_p^{\rm eff}$ being understood as an
effective number of modes.

\end{document}